\documentclass[1p]{elsarticle}

\usepackage{lineno,hyperref,natbib}
\modulolinenumbers[1]

\usepackage{xr}
\usepackage{bm}
\usepackage{soul}
\usepackage{float}

\usepackage{amsmath,amssymb,latexsym}
\usepackage{booktabs} % To thicken table lines
\usepackage[table]{xcolor} % Color in tables
\usepackage{array} % Align content of tables
\usepackage{rotating} % Rotates tables
\usepackage{pdfpages} 
\newcolumntype{P}[1]{>{\centering\arraybackslash}p{#1}}

\biboptions{sort&compress}

\begin{document}

\begin{frontmatter}

%\title{Assessing exponential random graph models for EEG brain networks}
\title{A statistical model for brain networks inferred from large-scale electrophysiological signals}

\author[mymainaddress,mysecondaryaddress]{Catalina Obando}
\author[mymainaddress,mysecondaryaddress]{Fabrizio De Vico Fallani\corref{mycorrespondingauthor}}
\cortext[mycorrespondingauthor]
{
Corresponding author\\ 
\\
ICM, Hopital Pitie-Salpetriere\\
75013 Paris, France\\
Tel +33(0)157274294\\
Email fabrizio.devicofallani@gmail.com
}

\address[mymainaddress]{Inria Paris, Aramis project-team, 75013, Paris, France}
\address[mysecondaryaddress]{Sorbonne Universites, UPMC Univ Paris 06, Inserm, CNRS, Institut du cerveau et la moelle (ICM), Hopital Pitie-Salpetriere, 75013, Paris, France}

\begin{abstract}
\small{
Network science has been extensively developed to characterize structural properties of complex systems, including brain networks inferred from neuroimaging data. 

As a result of the inference process, networks estimated from experimentally obtained biological data, represent one instance of a larger number of realizations with similar intrinsic topology. A modeling approach is therefore needed to support statistical inference on the bottom-up local connectivity mechanisms influencing the formation of the estimated brain networks.

Here, we adopted a statistical model based on exponential random graphs (ERGM) to reproduce brain networks, or connectomes, estimated by spectral coherence between high-density electroencephalographic (EEG) signals. ERGMs are made up by different local graph metrics whereas the parameters weight the respective contribution in explaining the observed network.
We validated this approach in a dataset of $N=108$ healthy subjects during eyes-open (EO) and eyes-closed (EC) resting-state conditions. 

Results showed that the tendency to form triangles and stars, reflecting clustering and node centrality, better explained the global properties of the EEG connectomes as compared to other combinations of graph metrics.
In particular, the synthetic networks generated by this model configuration replicated the characteristic differences found in the real brain networks, with EO eliciting significantly higher segregation in the $alpha$ frequency band ($8-13$ Hz) as compared to EC.
Furthermore, the fitted ERGM parameter values provided complementary information showing that clustering connections are significantly more represented from EC to EO in the \textit{alpha} range, but also in the \textit{beta} band ($14-29$ Hz), which is known to play a crucial role in cortical processing of visual input and externally oriented attention.

Taken together, these findings support the current view of the brain functional segregation and integration in terms of modules and hubs, and provide a statistical approach to extract new information on the (re)organizational mechanisms in healthy and diseased brains.
}
\end{abstract}

\begin{keyword}
\texttt{Exponential random graph models \sep Brain connectivity \sep Graph theory \sep EEG \sep Resting-states}
\end{keyword}

\end{frontmatter}

\newpage
\section{Introduction}
The study of the human brain at rest provides precious information that is predictive of intrinsic functioning, cognition, as well as pathology \citep{raichle_default_2001}.
In the last decade, graph theoretic approaches have described the topological structure of resting-state connectomes derived from different neuroimaging techniques, such as functional magnetic resonance imaging (fMRI) or magneto (MEG) and electroencephalography (EEG).

These estimated connectomes, or brain networks, tend to exhibit similar organizational properties including small-worldness, cost-efficiency, modularity and node centrality \citep{bullmore_complex_2009}, as well as characteristic dependence from the anatomical backbone connectivity \citep{honey_predicting_2009, deco_resting-state_2013, park_structural_2013} and genetic factors  \citep{fornito_genetic_2011}.
Furthermore, they show potentially clinical relevance as demonstrated by the recent development of network-based diagnostics of consciousness \citep{achard_hubs_2012, chennu_spectral_2014}, Alzheimer's disease \citep{tijms_alzheimers_2013}, stroke recovery \citep{grefkes_reorganization_2011}, and schizofrenia \citep{lynall_functional_2010}.
In this sense, quantifying topological properties of intrinsic functional connectomes by means of graph theory has enriched our understanding of the structure of functional brain connectivity maps \citep{stam_functional_2004,bullmore_complex_2009,rubinov_complex_2010,stam_organization_2012}.
Nevertheless, these results refer to a descriptive analysis of the observed brain network, which is only one instance of several alternatives with similar structural features. 
This is especially true for functional networks inferred from empirically obtained data, where the edges (or links) are noisy estimates of the true connectivity and thresholding is often adopted to filter the relevant interactions between the system units \citep{tumminello_tool_2005,vidal_interactome_2011, de_vico_fallani_graph_2014}.

Statistical models are therefore needed to reflect the uncertainty associated with a given observation, to permit inference about the relative occurrence of specific local structures, and to relate local-level processes to global-level properties \citep{goldenberg_survey_2009}.
A first approach consists in generating synthetic random networks that preserve some observed properties, such as the degree distribution or the random walk distribution, and then contrasting the values of the graph indices obtained in these synthetic networks with those extracted from the estimated connectomes \citep{rubinov_complex_2010}.
While these methods often provide appropriate null models, and can improve the identification of relevant network properties \citep{milo_network_2002,garlaschelli_patterns_2004, humphries_network_2008}, they do not inform on the organizational mechanisms modeling the whole network formation \citep{barabasi_evolution_2002,newman_random_2002}.
Alternative approaches consider probabilistic growth models such as those based on spatial distances between nodes \citep{barthelemy_spatial_2011}. Interesting results have been achieved in identifying some basic connectivity rules reproducing both structural and functional brain networks \citep{vertes_simple_2012,betzel_generative_2016}. However, these methods suffer from the rough approximation (e.g., Euclidean) of the actual spatial distance between nodes, and moreover, they do not indicate if the identified local mechanisms are either necessary or sufficient as descriptors of the global network structure.

To support inference on the processes influencing the formation of network structure, statistical models have been conceived to consider the set of all possible alternative networks weighted on their similarity to the observed one \citep{goldenberg_survey_2009}. 
Among others, exponential random graph model (ERGMs) represent a flexible category allowing to simultaneously assess the role of specific graph features on the formation of the entire network.
These models have been first proposed as an extension of the triad model defined by \cite{frank_markov_1986} to characterize Markov graphs \citep{frank_statistical_1991,wasserman_logit_1996}  and have been widely developed to understand how simple interaction rules, such as transitivity, could give rise to the complex network of social contacts \citep{handcock_statistical_2002,hunter_inference_2006,goodreau_advances_2007,hunter_curved_2007,rinaldo_geometry_2009, robins_closure_2009,goodreau_birds_2009,wang_exponential_2013,niekamp_sexual_2013}.

Recently, the use of ERGM has been proved to successfully model imaging connectomes derived respectively from spontaneous fMRI activity \citep{simpson_exponential_2011} and diffusion tensor imaging (DTI) \citep{sinke_bayesian_2016}. 
%In those studies, authors were able to statistically reproduce the constructed brain networks by fitting ERGMs with three main graph features. The first one, the connection density, had an obvious interpretation. The other two, nominally the geometrically weighted edgewise shared partner ($GW_{E}$) and non-edgewise shared partner ($GW_{N}$), have been instead associated to the global-efficiency and clustering (local-efficiency), respectively.
Despite its potential, the use of ERGM in network neuroscience is still in its infancy and more evidence is needed to better elucidate its applicability to connectomes inferred from other types of neuroimaging data and across different experimental conditions.
In addition, many methodological issues remain unanswered, like for example the relationships between the graph metrics included in the ERGM and the graph indices used to describe the topology of the observed connectomes. 

To address the above issues, we proposed and evaluated several ERGM configurations based on the combination of different local connectivity structures (i.e., graph metrics). Specifically, we modeled brain networks estimated from high-density EEG signals in a group of healthy individuals during eyes-open (EO) and eyes-closed (EC) resting-states. 
Our goal was to identify the best ERGM configuration reproducing EEG-derived connectomes in terms of functional integration and segregation, and to evaluate the ability of the estimated ERGM parameters in providing new information discriminating between EO and EC conditions.

\section{Methods}
\label{subsec:methods}

\subsection{EEG data and brain network construction}
We used high-density EEG signals freely available from the online PhysioNet BCI database \citep{goldberger_physiobank_2000,schalk_bci2000:_2004}. EEG data consisted of one minute resting-state with eyes-open (EO) and one minute resting-state with eyes-closed (EC) recorded from $56$ electrodes in $108$ healthy subjects. EEG signals were recorded with an original sampling rate of 160Hz. All the EEG signals were referenced to the mean signal gathered from electrodes on the ear lobes. We subsequently downsampled the EEG signals to 100Hz after applying a proper anti-aliasing low-pass filter. The electrode positions on the scalp followed the standard 10-10 montage.

We used the spectral coherence \citep{carter_coherence_1987} to measure functional connectivity (FC) between EEG signals of sensors $i$ and $j$ at a specific frequency band $f$ as follows:
\begin{equation}
w_{ij}(f)=\frac{\mid S_{ij}(f)\mid ^2}{S_{ii}(f)S_{jj}(f)}
\label{Eq:1}
\end{equation}
Where $S_{ij}$ is the cross-spectrum between $i$ and $j$, and $S_{ii}$ and $S_{jj}$ the autospectra of $i$ and $j$ respectively. 
Specifically, we computed cross- and auto-spectra by means of the Welch's averaged modified periodogram with a sliding Hanning window of $1$s and $0.5$ seconds of overlap. The number of FFT points was set to $100$ for a frequency resolution
of $1$Hz.
As a result, we obtained for each subject a connectivity matrix $W(f)$ of size $56\times 56$ where the entry $w_{ij}(f)$ contains the value of the spectral coherence between the EEG signals of sensors $i$ and $j$ at the frequency $f$.

We then averaged the connectivity matrices within the characteristic frequency bands \textit{theta} ($4-7$ Hz), \textit{alpha} ($8-13$ Hz), \textit{beta} ($14-29$ Hz), \textit{gamma} ($30-40$ Hz). These matrices constituted our raw brain networks whose nodes corresponded to the EEG sensors ($n=56$) and links corresponded to the $w_{ij}$ values.
Finally, we thresholded the values in the connectivity matrices to retain the strongest links in each brain network. Specifically, we adopted an objective criterion, i.e., the efficiency cost optimization, to filter and binarize a number of links such that the final average node degree $k=3$ \citep{de_vico_fallani_topological_2017}. We also considered $k=1, 2, 4, 5$ to evaluate the main brain network properties around the representative threshold $k=3$ . The resulting sparse brain networks, or graphs, were represented by adjacency matrices $A$, where each entry indicates the presence $a_{ij}=1$ or absence $a_{ij}=0$ of a link between nodes $i$ and $j$.

\subsection{Graph indices}
\label{subsec: comparison}
% Graph Analysis: Topological quantities
% - first introduce the graph quantities
% - compare quantities between EC and EO
We evaluated the global structure of brain networks by measuring graph indices at large-scale topological scales.
We focused on well-known properties of brain networks such as optimal balance between integration and segregation of information \citep{tononi_measure_1994,bassett_small-world_2006,bullmore_complex_2009}. Integration is the tendency of the network to favor distributed connectivity among remote brain areas; conversely, segregation is the tendency of the network to maintain connectivity within specialized groups of brain areas \citep{friston_functional_2011}.

In graph theory integration has been typically quantified by the global-efficiency $E_g$ and by the characteristic path length $L$: 

\begin{equation}
\label{Eq:global}
\begin{aligned}
E_g&=\frac{1}{n(n-1)}\sum\limits_{i,j=1,i\neq j}^n \frac{1}{d_{ij} }\\
L&=\frac{1}{n(n-1)}\sum\limits_{i,j=1, i\neq j}^n d_{ij}
\end{aligned} 
\end{equation}

Where $d_{ij}$ is the distance, or the length of the shortest path, between nodes $i$ and $j$ \citep{watts_collective_1998,latora_efficient_2001}.

Segregation is typically measured by means of the local-efficiency $E_l$ and by the clustering coefficient $C$:

\begin{equation}
\label{Eq:local}
\begin{aligned}
E_{l}&=\frac{1}{n}\sum\limits_{i=1}^n E_g(G_i)\\
C&=\frac{1}{n}\sum\limits_{i=1}^n \frac{2t_i}{k_i(k_i-1) } \\
\end{aligned} 
\end{equation}

where $G_i$ is the subgraph formed by the nodes connected to $i$; $t_i$ is the number of triangles around node $i$, and $k_i$ is the degree of node $i$ \citep{watts_collective_1998,latora_efficient_2001}.

In addition, we evaluated the strength of division of a network into modules by measuring the modularity $Q$:

\begin{equation}
\label{Eq:Modularity}
\begin{aligned}
Q&= \frac{1}{l} \sum\limits_{i,j=1}^n \left( A_{ij}- \frac{k_ik_j}{l}\right)\delta_{m_i,m_j}
\end{aligned} 
\end{equation}
where $l = \sum\limits_{i,j=1}^n A_{ij}$ is the number of edges, $m_i$ is the module containing node $i$, and $\delta_{m_i,m_j}=1$ if $m_i=m_j$ and $0$ otherwise. We used the Walktrap algorithm to generate a sequence of community partitions \citep{pons_computing_2005} and we selected the one that maximized $Q$ according to the standard algorithm proposed in \cite{newman_modularity_2006}. Modularity can be seen as a compact measure of the integration and segregation of a network, as it measures the propensity to form dense connections between nodes within modules (i.e., segregation) but sparse connections between nodes in different modules (i.e., inverse of integration).

\subsection{Exponential random graph model (ERGM)}
\label{subsec:generalERGM} 

Let $G$ be a graph in a set $\mathcal{G}$ of possible network realizations, $g=[g_1, g_2,...,g_r]$ be a vector of graph statistics, or metrics, and $g^*=[g_1^*, g_2^*,...,g_r^*]$ the values of these metrics measured over $G$. 
Then, we can statistically model $G$ by defining a probability distribution $P(G)$ over $\mathcal{G}$ such that the following conditions are satisfied:

\begin{equation}\label{eq:cnstr1}
\sum\limits_{G\in \mathcal{G}}P(G)= 1
\end{equation}
\begin{equation}\label{eq:cnstr2}
\langle g_i\rangle =\sum\limits_{G\in \mathcal{G}}g_i(G)P(G)=g_i^*,\quad i=\{1,2,...,r\}
\end{equation}

where $\langle g_i\rangle$ is the expected value of the $i-th$ graph metric over $\mathcal{G}$.

By maximizing the Gibbs entropy of $P(G)$ constrained to the above conditions, the probability distribution reads as:

\begin{equation}
P(G)=\frac{e^{H(G)}}{Z}\label{Eq:GeranlERGM}
\end{equation}

where $H(G)=\sum\limits_{i=1}^{r}\theta_i g_i(G)$ is the graph Hamiltonian, $\theta_i$ is the $i-th$ model parameter to be estimated and $Z = \sum\limits_{G\in\mathcal{G}}e^{H(G)}$ is the so-called partition function \citep{newman_networks:_2010}.
The estimated value of a parameter $\theta_i$ indicates the change in the (log-odds) likelihood of an edge for a unit change in graph metric $g_i$. If the estimated value of $\theta_i$ is large and positive, the associated graph metric $g_i$ plays an important role in explaining the topology of $G$ more than would expected by chance. Notice that here chance corresponds to randomly choosing a network from the space $\mathcal{G}$. 
If instead the estimated value of $\theta_i$ is negative and large then $g_i$ still plays an important role in explaining the topology of $G$ but is less prevalent than expected by chance \citep{robins_introduction_2007}.

In general, the fact that the space $\mathcal{G}$ can be very large even for relatively small $n$, as well as the inclusion of graph metrics that are not simple linear combinations of $G_{ij}$, makes in practice impossible to derive analytically the model parameters vector $\boldsymbol{\theta}=\left[ \theta_1,\theta_2,...,\theta_r \right]$ \citep{frank_markov_1986, hunter_inference_2006}.

Numerical methods, such as Markov chain Monte Carlo (MCMC) approximations of the maximum likelihood estimators (MLE) of the model parameters vector $\boldsymbol{\theta}$ are typically adopted to circumvent this issue \citep{snijders_markov_2002}.

\subsubsection{Model construction and implementation}
\label{subsec:bcERGM}

\begin{figure}[!ht]
\centering
\includegraphics[width=70.0mm]{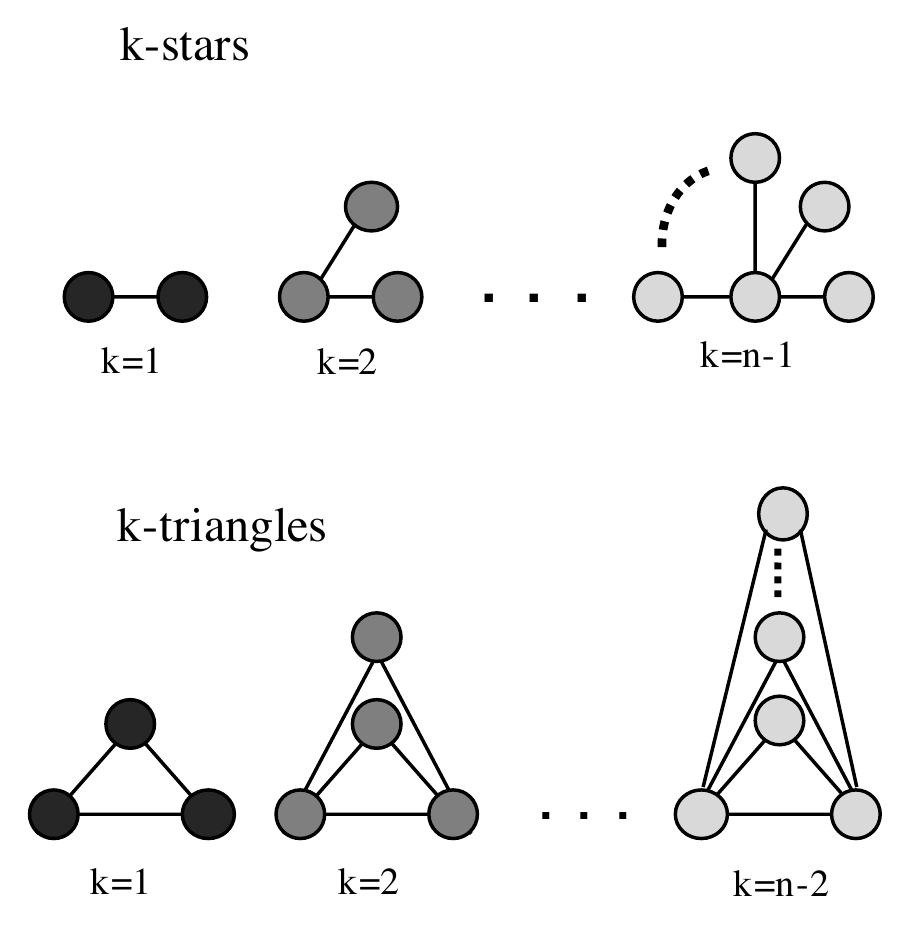}
\caption{Graphical representation of $k-$stars and $k-$triangles.}
%    \caption{ Graphical representation of graph metrics used in the proposed models. $GW_K$ is a geometrically weighted sequence of $k$-stars which are closely related to the concept of hubs. $GW_E$ is a geometrically weighted sequence of the shared pattern distribution $S$ when considering only connected dyads; $GW_N$ is a geometrically weighted sequence of the shared pattern distribution $S$ when considering only non-connected dyads and; $GW_D$  is a geometrically weighted sequence of the shared pattern distribution $S$ when considering any dyad (connected and non-connected). $GW_E$ is closely related to the number of triangles; $GW_N$ to paths of length $2$ and; $GW_D$ satisfies $GW_D = GW_E + GW_N $. The ratio parameter of the geometrically weighted sequence $\tau$ gives less weight to higher order structures (i.e., high $k$), this is depicted in the picture with the gradient grey scale where darker colors are associated to higher weights and lighter colors to lower weights.}
    \label{fig:Figure0}
\end{figure}

%introduce metrics typically used and validated in social network sciences
We considered graph metrics reflecting basic properties of complex systems such as hub propensity and transitivity in the network \citep{amaral_classes_2000,wang_complex_2003, bassett_small-world_2006}.
Specifically, we focused on $k$-stars to model highly connected nodes (hubs) and $k$-triangles to model transitivity, where $k$ refers to the order of the structures as illustrated in Fig. \ref{fig:Figure0}. 

In general, this leads to a large number of model parameters to be estimated, i.e., $n-1$ for $k$-stars and $n-2$ for $k$-triangles.
To avoid consequent degeneracy issues in the ERGM estimation, we adopted a compact specification for those metrics that combines them in an alternating geometric sequence \citep{snijders_new_2006,hunter_inference_2006}. 

Because $k$-stars are related to the node degree distribution $D$ \citep{hunter_curved_2007}, we used the \textit{geometrically weighted degree distribution} $GW_K$ as a graph metric to characterize hub propensity:

\begin{equation}
GW_K=e^{2\tau} \sum\limits_{i=1}^{n-1} ((1 - e^{-\tau} )^i -1 +ie^{-\tau }) D_i
\label{Eq: GWKSP}
\end{equation}

where $\tau>0$ is a \textit{ratio parameter} to penalize nodes with extremely high node degrees. 

Similarly, because $k$-triangles are related to the \textit{shared pattern distribution} $S$, we used the \textit{geometrically weighted edgewise shared partner distribution} to characterize transitivity:

\begin{equation}
GW_E=e^\tau \sum\limits_{i=1}^{n-2} (1 - (1 - e^{-\tau} )^i ) S_i
\label{Eq: sharedPttrn}
\end{equation}

where the element $S_i$ is the number of dyads that are directly connected and that have exactly $i$ neighbors in common. 

In addition, complementary metrics have been defined based on the \textit{shared partner distribution}:

\begin{itemize}

\item[$GW_N$:] \textit{geometrically weighted non-edgewise shared partner distribution} given by Eq. \ref{Eq: sharedPttrn}, with $S_i$ considering exclusively dyads that are not connected,

\item[$GW_D$:] \textit{geometrically weighted dyadwise shared partner distribution} given by Eq. \ref{Eq: sharedPttrn}, with $S_i$ considering any dyad, connected or not.

\end{itemize}

The above specifications yield particular ERGMs that belong to the so-called curved exponential family \citep{hunter_curved_2007} and that have been extensively used in social science \citep{robins_recent_2007,goodreau_advances_2007,lusher_exponential_2012}. 

We constructed different ERGMs configurations by including these graph metrics as illustrated in Table \ref{table:testGeo}. For the sake of simplicity, we only considered combinations of two graph metrics at most, except in one case where we also included the number of edges as a further metric \citep{simpson_exponential_2011, sinke_bayesian_2016}.

%application/implentation
We tested the different configurations by fitting ERGM to brain networks in each single subject ($N=108$), frequency band ($theta$, $alpha$, $beta$, $gamma$), and condition (EO, EC).
To fit ERGMs we used a Markov chain Monte Carlo (MCMC) algorithm (Gibbs sampler) that samples networks from an exponential graph distribution. Specifically, we set the initial values of the model parameters $\boldsymbol{\theta^0}$ by means of a maximum pseudo-likelihood estimation (MPLE) \citep{snijders_markov_2002, van_duijn_framework_2009}. 
Then, we adopted the Fisher's scoring method to update the model parameters $\boldsymbol{\theta}$ until they converge to the approximated maximum likelihood estimators (MLEs) $\hat{\theta}$ \citep{hunter_inference_2006}. 
As we used curved exponential random graph models, the ratio parameters $\tau$ were not fixed but were estimated as well.

Eventually, for each fitted ERGM configuration we generated $100$ synthetic networks in order to obtain appropriate confidence intervals.

\subsubsection{Goodness of fit}
\label{subsec:GOF}
First, we used the Akaike information criterion (AIC) to evaluate relative quality of the ERGMs' fit by taking into account the maximum value of the likelihood function and the number of model parameters \citep{akaike_information_1998}.
% second criterion relative error with respect to eglob and eloc

We also adopted a different approach to assess the absolute quality of the fit by comparing the synthetic networks generated by the estimated ERGMs and the observed brain networks.
Specifically, we defined the following score based on the integration and segregation properties of networks:

\begin{equation}
\delta(E_g, E_l) = max\left( \mid\eta_{E_g}\mid,\mid\eta_{E_l}\mid \right)
\label{Eq: RE}
\end{equation}
Where $\eta_{E_g}, \eta_{E_l}$ are the relative errors between the mean values of global/local efficiency of the simulated networks and the value of the observed brain network. By selecting the maximum absolute error we then considered the worse case similar to what have been proposed in \cite{betzel_generative_2016}.
% Model Adequacy
Based on the above criteria, we selected the best model, which minimizes the AIC and $\delta$ mean values. 
To validate the model adequacy (Eq.\ref{eq:cnstr2}) we computed the Z-scores between the graph metrics' values of brain networks and synthetic networks. 

Furthermore, we cross-validated the best model configuration by evaluating the synthetic networks' fitness to graph indices that were explicitly included neither in the ERGM nor in the model selection. We computed Pearson's correlation coefficient between the values of the characteristic path length ($L$), clustering coefficient ($C$) and modularity ($Q$) extracted from the observed brain networks and the mean values obtained from the corresponding simulated networks.
In addition, we used the Mirkin index (MI) \citep{meila_comparing_2007} to evaluate the similarity between the community partitions of the observed networks and the consensus partitions of the corresponding synthetic networks.

\subsection{Statistical group analysis}

We assessed the statistical differences between the values of the graph indices extracted from the brain networks in EO and EC resting state conditions. 
We also computed between-condition differences using the synthetic networks fitted by the best ERGMs. In this case, we considered the mean values of the graph indices in order to have one value corresponding to one brain network. 
Eventually, we computed the statistical differences between the values of the ERGM parameters in the EO and EC condition in order to assess their potential to provide complementary information as compared to standard graph analysis.
For each comparison, we used a non-parametric permutation t-test and we fixed a statistical threshold of $\alpha=0.001$ and $100000$ permutations.

\section{Results}\label{sec:Results}

\subsection{Characteristic functional segregation of EEG resting-state networks}
\begin{figure}[!ht]
\centering
\includegraphics[width=90.0mm]{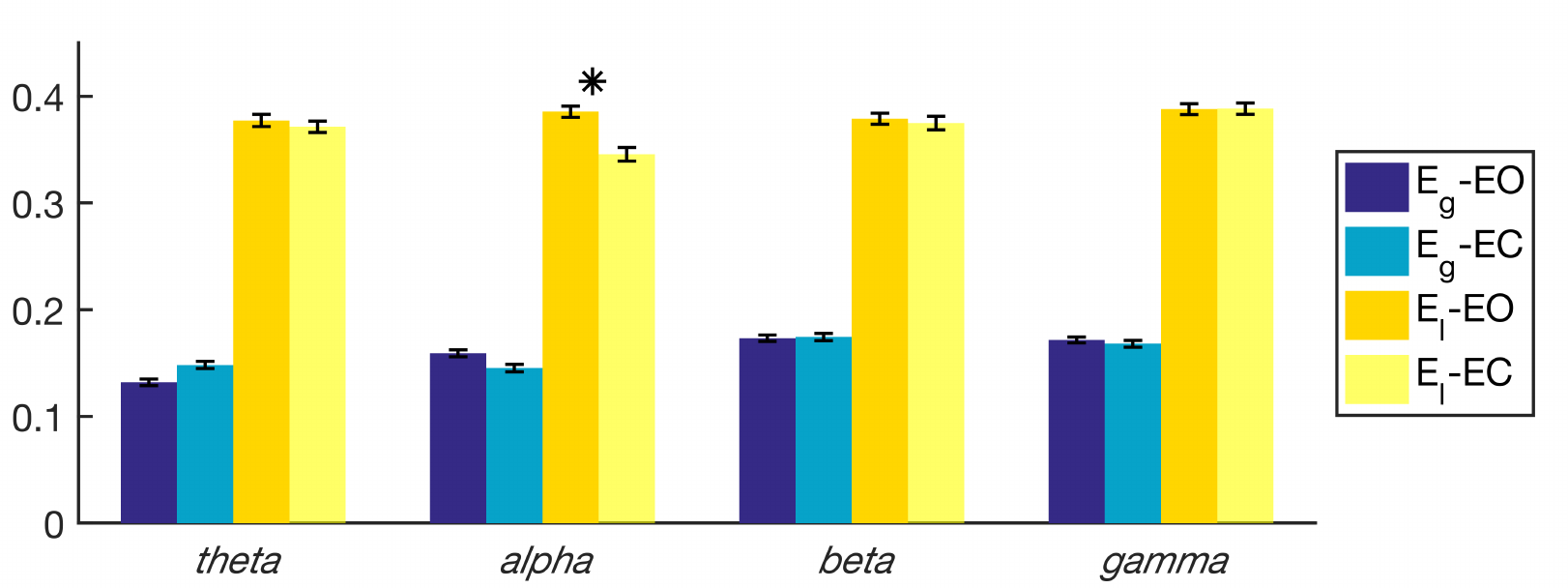}
    \caption{Median values and standard errors of global- and local-efficiency measured from EEG brain networks across 108 subjects in eyes-open (EO) and eyes-closed (EC) resting-states. \newline 
$\ast$ p-value $< 0.001$.}
    \label{fig:Figure1}
\end{figure}

The group-analysis revealed a significant increase of the local-efficiency in EO, as compared to EC, for the $alpha$ band ($T=3.529, p=0.0007$, Fig. \ref{fig:Figure1}). 
We also reported a significant increment ($T=3.557, p=0.0007$) for the modularity in the $alpha$ band, while no other statistically significant differences were observed in the other frequency bands, graph indices or metrics (Table \href{TableS3.pdf}{S3}).

These differences were obtained for brain networks thresholded with an average node degree $k=3$ according to the ECO criterion \citep{de_vico_fallani_topological_2017}. 
Because we reported a similar increase of functional segregation (local-efficiency) in the $alpha$ band for $k=5$ (Fig. \href{FigureS1.pdf}{S1}), we adopted here $k=3$ as a representative threshold. More details on the analysis for $k=5$ can be found in the \href{Supp_Text.pdf}{Supplementary text}.

In term of existing relationships between graph indices and ERGM metrics, we could not establish univocal associations between $E_g$ and $E_l$ values and the metrics' values used in the ERGMs (Table \href{TableS1.pdf}{S1}). This was especially true for the global-efficiency, which exhibited significantly high correlations with all the other graph metrics (Spearman's $\mid R \mid >0.43, p<10^{-39}$).

\subsection{Triangles and stars as fundamental constituents of functional brain networks}
\begin{figure}[!ht]
\centering
\includegraphics[width=135.0mm]{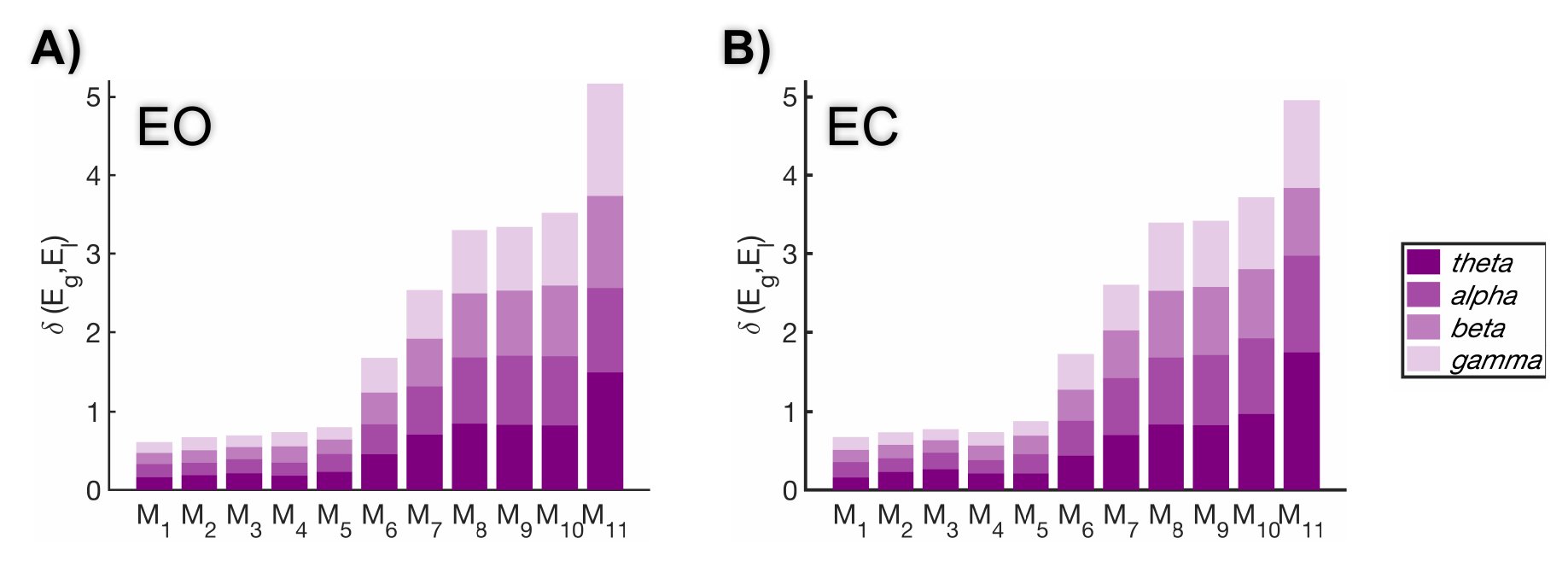}
    \caption{Absolute quality of ERG models' fit. Colored bars show the group-averaged cumulative errors $\delta(E_g,E_l)$ in terms of relative of global- and local-efficiency across frequency bands. Model configurations are listed on the x-axis. Panel A) illustrates values for eyes-open resting-state (EO); panel B) shows the error values for eyes-closed resting-state (EC).}
    \label{fig:Figure2}
\end{figure}

All the proposed ERGM configurations exhibited a relatively good fitting in terms of AIC, except for $M_{11}$ (Fig. \href{FigureS2.pdf}{S2}). Notably, the latter was the only configuration where the number of edges was considered as a model parameter and not as a constraint. 
$M_1$ gave the lowest $\delta(E_g,E_l)$ scores as compared to other configurations in both EO and EC conditions (Fig. \ref{fig:Figure2}). Notably, the configurations giving lower $\delta(E_g,E_l)$ scores included, directly or indirectly, the metric $GW_E$, with the exception of $M_{11}$. 

We selected $M_1$ as a potentially good candidate to model EEG-derived brain networks. According to this model configuration the mass probability density reads $P(G)= Z^{-1}exp \left\{ \theta_1GW_E+\theta_2GW_K \right\}$.
The group-median values of the estimated parameters ($\theta_1$ and $\theta_2$) were all positive and larger than $1$ in each band and condition (Table \ref{table2}). 
This means that the likelihood of an edge to exist in a simulated network is larger if that edge is part of a triangle ($GW_E$) or of a star ($GW_K$) and that these connectivity structures are statistically relevant for the brain network formation.

\begin{figure}[!ht]
\centering
\includegraphics[width=120.0mm]{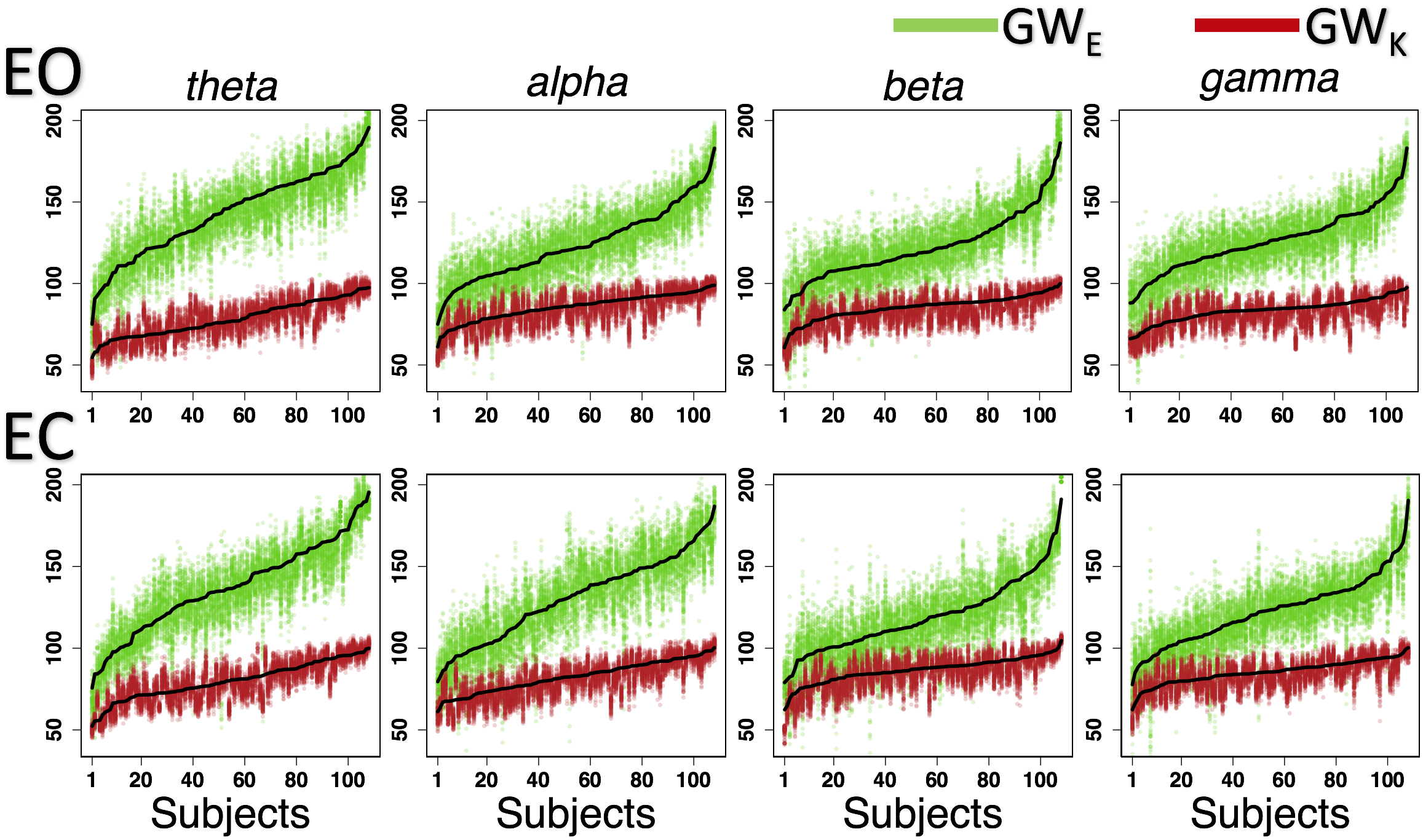}
    \caption{Adequacy of model configuration $M_1$. Green and red dots represent respectively the values of the \textit{geometrically weighted edgewise shared pattern distribution} ($GW_E$) and \textit{geometrically weighted degree distribution} ($GW_K$) measured in simulated networks. Black dots lines indicate the values measured in the observed brain networks.}
    \label{fig:Figure3}
\end{figure}

\begin{figure}[!ht]
\centering
\includegraphics[width=110.0mm]{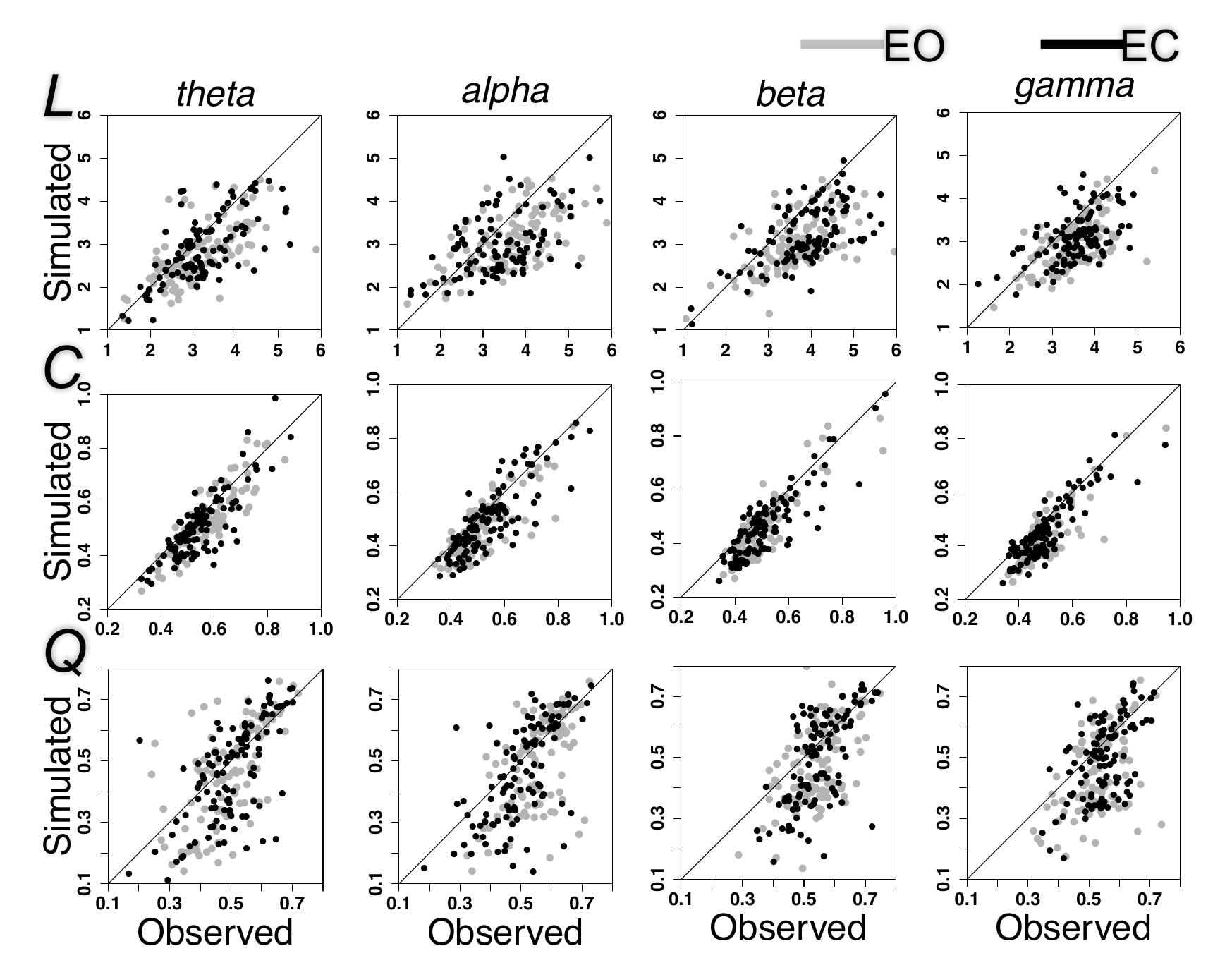}
    \caption{Cross-validation for model configuration $M_1$. Scatter plots show the values of the graph indices measured in the observed brain networks (x-axis) against the mean values obtained from synthetic networks (y-axis). Three graph indices were considered: characteristic path length ($L$), clustering coefficient ($C$) and modularity ($Q$). Grey dots correspond to eyes-open resting-states (EO); black dots correspond to eyes-closed resting-states (EC).}
    \label{fig:Figure4}
\end{figure}

Overall, the $GW_E$ and $GW_K$ values of the synthetic networks generated by $M_1$ were not significantly different from those of the observed brain networks (Fig. \ref{fig:Figure3}). This was true in every subject for $GW_E$ ($Z<2.58$, $p>0.01$) and in at least the $94\%$ of the subjects for $GW_K$ ($Z<2.58$, $p>0.01$).
Furthermore, the values of the characteristic path length ($L$), clustering coefficient ($C$) and modularity ($Q$), extracted from synthetic networks were significantly correlated (Pearson's $R>0.44, p<10^{-6}$) with those of the brain networks in each frequency band (Fig. \ref{fig:Figure4}, Table \href{TableS2.pdf}{S2}). In addition, synthetic networks exhibited a similar community partition compared to individual brain networks, as revealed by the low Mirkin index values (MI$<0.21$) (Fig. \href{FigureS3.pdf}{S3}). These results confirmed that $M_1$ adequately models the obtained EEG brain networks.

\subsection{Simulating network differences between absence or presence of visual input}
\label{subsec:GroupDifferences}

\begin{figure}[!ht]
\centering
\includegraphics[width=120.0mm]{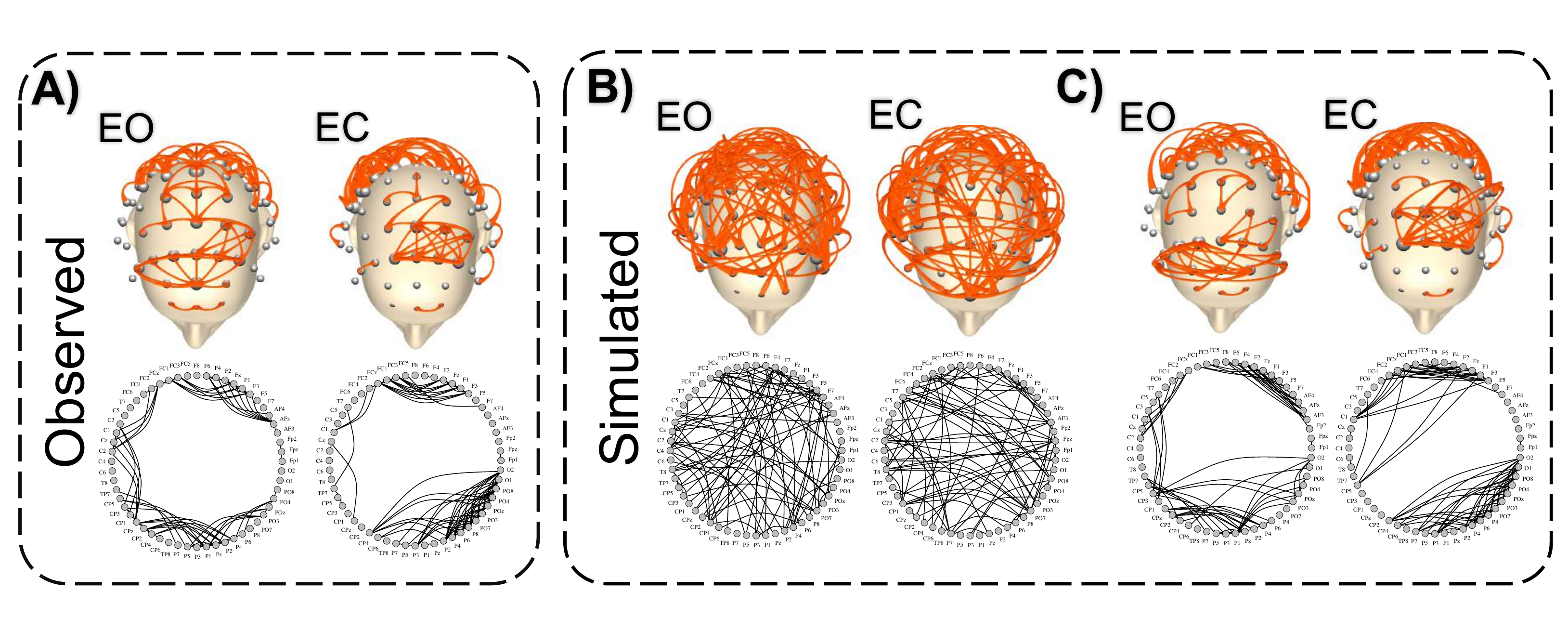}
    \caption{Brain networks and synthetic networks for a representative subject. Panel A): brain network in the $alpha$ band for the eyes-open (EO) and eyes-closed (EC) resting-state. Panel B): one instance of the corresponding synthetic networks generated by model configuration $M_1$. Panel C): because node labels are not preserved in the simulated networks, we re-assigned them virtually by using the Frank-Wolfe algorithm \citep{vogelstein_fast_2011}, which optimizes the graph matching with the observed brain network.  
In the upper part of the figure, nodes correspond to EEG electrodes, whose position follows a standard $10-10$ montage. In the bottom part, the nodes are arranged into a circle.  }
    \label{fig:Figure5}
\end{figure}

Fig. \ref{fig:Figure5} illustrates the brain networks for a representative subject in the $alpha$ band along with corresponding synthetic networks generated by $M_1$. In both EO and EC conditions, simulated networks and brain networks share similar topological structures characterized by diffused regularity and more concentrated connectivity in parietal and occipital regions.

The group-analysis over the synthetic networks revealed the ability of $M_1$ to capture not only individual properties of brain networks, but also the main observed difference between EC and EO resting-states, reflecting respectively absence and presence of visual input. 
Similarly to observed brain networks, we obtained, for simulated networks, a marginal significant increase of the local-efficiency from EC to EO, in the $alpha$ band ($T=3.168, p=0.002$). No other significant differences were reported in any other band or graph index/metric (Table \href{TableS3.pdf}{S3}).

Finally, by looking at the values of the estimated parameters, we observed that $\theta_1$ values were significantly larger in EO compared to EC for both $alpha$ ($T=3.746, p=0.0002$) and $beta$ ($Z=1.514, p=0.0009$) frequency bands, while no significant differences were found for $\theta_2$ values (Table \ref{table2}). 

\section{Discussion}
\label{sec:Discussion}

In the last years, the use of statistical methods to infer the structure of complex systems has gained increasing interest \citep{simpson_exponential_2011,guimera_network_2013, martin_structural_2016,hric_network_2016}.
Beyond the descriptive characterization of networks, statistical network models (SNM) aim to statistically assess the local connectivity processes involved in the global structure formation \citep{goldenberg_survey_2009}.
This is a crucial advance with respect to standard descriptive approaches because imaging connectomes, as other biological networks, are often inferred from experimentally obtained data and therefore the estimated edges can suffer from statistical noise and uncertainty \citep{craddock_imaging_2013}.

In our study, we used ERGMs to identify the local connectivity structures that statistically form the intrinsic synchronization of large-scale electrophysiological activities. 
This model formulation has the advantage of statistically infer the probability of edge formation accounting for highly dependent configurations, such as transitivity structures, something lacking for example in the Bernoulli model. Furthermore, it is possible to include, in theory, graph metrics measuring global and local properties and discriminating node and edges attributes, such as homophily effects.
In addition, it generalizes well-known networks models such as the stochastic block-model, where a block structure is imposed by including the count of edges between groups of nodes as a model metric \citep{holland_stochastic_1983}.

Here, results showed that the tendency to form triangles ($GW_E$) and stars ($GW_K$) were sufficient to statistically reproduce the main properties of the EEG brain networks, such as functional integration and segregation, measured by means of global- $E_g$ and local-efficiency $E_l$ (Table \href{TableS3.pdf}{S3}).
Our findings partially deviate from previous studies having adopted an ERGM to model fMRI and DTI brain networks, where $GW_E$ and the geometrically weighted non-edgewise shared partner $GW_N$ were selected under the assumption that these could be related respectively to local- and global-efficiency \citep{simpson_exponential_2011, sinke_bayesian_2016}. 
However, here we showed that an univocal relationship between the ERGM graph indices and the metrics used to describe the EEG connectomes could not be statistically established (Table \href{TableS1.pdf}{S1}).
While the propensity to form triangles ($GW_E$) can lead to cohesive clustering in the network ($E_l$), the propensity to form redundant paths of length $2$ (i.e., $GW_N$) is not clearly related to the formation of short paths between nodes ($E_g$) \citep{snijders_new_2006}. 
Thus, while in general a good fit can be achieved including $GW_N$ in the ERGM, the subsequent interpretation in terms of brain functional integration appears less straightforward.
Here, we showed that $GW_E$ together with the tendency to form stars ($GW_K$), rather gave the best fit in terms of local- and global-efficiency. Triangles and stars, giving rise to clustering and hubs, are fundamental building blocks of complex systems reflecting important mechanisms such as transitivity \citep{watts_collective_1998} and preferential attachment \citep{barabasi_emergence_1999}.
Notably, the existence of highly connected nodes is compatible with the presence of short paths (e.g., in a star graph the characteristic path length $L=2$). This supports the recent view of the brain functional integration where segregated modules exchange information through central hubs and not necessarily through shortest paths \citep{sporns_network_2013, deco_rethinking_2015}.

%Results of higher segregation in EO, motivate
In the cross-validation phase, the selected model configuration captured other important brain network properties as measured by the clustering coefficient $C$, the characteristic path length $L$, and the modularity $Q$ (Fig. \ref{fig:Figure4}). 
In terms of differences between conditions, the simulated networks gave a marginal significant increase ($p=0.002$) of $E_l$ in the $alpha$ band during EO as compared to EC, while, differently from observed brain networks, no significant differences were reported for the modularity $Q$ (Table \href{TableS3.pdf}{S3}).
The latter could be in part ascribed to the absence of specific metrics in the ERGM accounting for modularity. In such respect, stochastic block models, which explicitly force modular structures, could represent an interesting alternative to explore in the future \citep{karrer_stochastic_2011, pavlovic_stochastic_2014}.
Here, the increased $alpha$ local-efficiency suggests a modulation of augmented specialized information processing, from EC to EO, that is consistent with typical global power reduction and increased regional activity \citep{barry_eeg_2007}. Possible neural mechanisms explaining this effect have been associated to automatic gathering of non-specific information resulting from more interactions within the visual system \citep{yan_spontaneous_2009} and to shifts from interoceptive towards exteroceptive states \citep{marx_eyes_2004,bianciardi_modulation_2009,xu_different_2014}.

%Results of of the parameters in alpha and beta band only for GWE 
As a crucial result, we provided complementary information by inspecting the fitted ERGM parameters. The positive $\theta_1>1$ and $\theta_2>1$ values indicated that both $GW_E$ and $GW_K$ are fundamental connectivity features that emerge in brain networks more than expected by chance (Table \ref{table2}).
However, only $\theta_1$ values showed a significant difference (EO$>$EC) in the $alpha$ band, as well as in the $beta$ band (Table \ref{table2}), suggesting that the tendency to form triangles, rather than the tendency to form stars, is a discriminant feature of eyes-closed and eyes-closed modes.
More concentrated EEG activity among parieto-occipital areas has been largely documented in the $alpha$, but also in $beta$  band, the latter reflecting either cortical processing of visual input or externally oriented attention \citep{barry_eeg_2007, boytsova_eeg_2010}.
Notably, the role of the $beta$ band could be found neither when analyzing brain networks nor synthetic networks (Fig.\ref{fig:Figure1}, Table \href{TableS3.pdf}{S3}) and we speculate that this result specifically stems from the inherent ability of ERGMs to account for potential interaction between different graph metrics \citep{snijders_new_2006}.

\subsection{Methodological considerations}
We estimated EEG connectomes by means of spectral coherence. While this measure is known to suffer from possible volume conduction effects \citep{srinivasan_eeg_2007}, it has been also demonstrated that, probably due to this effect, it has the advantage of generating connectivity matrices highly consistent within and between subjects \citep{colclough_how_2016}.
In addition, spectral coherence is still one of the most used measures to infer functional connectivity in the electrophysiological literature of resting-states because of its simplicity and relatively intuitive interpretation. Thus, constructing EEG connectomes by means of spectral coherence allowed us to better contextualize the results obtained with ERGM from a neurophysiological perspective. Future studies will have to assess if and how different connectivity estimators affect the choice of the model parameters.

We used a density-based thresholding procedure to filter information in the EEG raw networks by retaining and binarizing the strongest edges.
Despite the consequent information loss, thresholding is often adopted to mitigate the uncertainty of the weakest edges, reduce the false positives, and facilitate the interpretation of the inferred network topology \citep{rubinov_complex_2010, de_vico_fallani_graph_2014}.

Selecting a binarizing threshold does have an impact on the topological structure of brain networks \citep{garrison_stability_2015}. Based on the optimization of fundamental properties of complex systems, i.e., efficiency and economy, the adopted thresholding criterion (ECO) leads to sparse networks, with an average node degree $k=3$, causing possible nodes to be disconnected. However, it has demonstrated empirically that the size of the resulting largest component typically contains more than the $60\%$ of the brain nodes, thus ensuring a sparse but meaningful network structure \citep{de_vico_fallani_topological_2017}.
In a separate analysis, we verified that the validity of the model and the characteristic between-condition differences observed in the $alpha$ band, were also globally preserved for $k=5$. (\href{Supp_Text.pdf}{Supplementary text}).

\section*{Author Contributions}
CO and FDVF both designed the study. CO performed the analysis and wrote the paper. FDVF wrote the paper.

\section*{Acknowledgments}
We are grateful to M. Chavez and J. Guillon for their useful comments and suggestions. 

\section*{Funding Statement}
This work has been partially supported by the French program ANR-10-IAIHU-06 and ANR-15-NEUC-0006-02.

\bibliography{CollectionCatalina}

\begin{thebibliography}{10}
\expandafter\ifx\csname url\endcsname\relax
  \def\url#1{\texttt{#1}}\fi
\expandafter\ifx\csname urlprefix\endcsname\relax\def\urlprefix{URL }\fi
\expandafter\ifx\csname href\endcsname\relax
  \def\href#1#2{#2} \def\path#1{#1}\fi

\bibitem{raichle_default_2001}
M.~E. Raichle, A.~M. MacLeod, A.~Z. Snyder, W.~J. Powers, D.~A. Gusnard, G.~L.
  Shulman, \href{http://www.pnas.org/content/98/2/676}{A default mode of brain
  function}, Proceedings of the National Academy of Sciences 98~(2) (2001)
  676--682.
\newblock \href {http://dx.doi.org/10.1073/pnas.98.2.676}
  {\path{doi:10.1073/pnas.98.2.676}}.
\newline\urlprefix\url{http://www.pnas.org/content/98/2/676}

\bibitem{bullmore_complex_2009}
E.~Bullmore, E.~Bullmore, O.~Sporns, O.~Sporns, Complex brain networks: graph
  theoretical analysis of structural and functional systems, Nat Rev Neurosci
  10~(maRcH) (2009) 186--198.
\newblock \href {http://dx.doi.org/10.1038/nrn2575}
  {\path{doi:10.1038/nrn2575}}.

\bibitem{honey_predicting_2009}
C.~J. Honey, O.~Sporns, L.~Cammoun, X.~Gigandet, J.~P. Thiran, R.~Meuli,
  P.~Hagmann, \href{http://www.pnas.org/content/106/6/2035}{Predicting human
  resting-state functional connectivity from structural connectivity},
  Proceedings of the National Academy of Sciences 106~(6) (2009) 2035--2040.
\newblock \href {http://dx.doi.org/10.1073/pnas.0811168106}
  {\path{doi:10.1073/pnas.0811168106}}.
\newline\urlprefix\url{http://www.pnas.org/content/106/6/2035}

\bibitem{deco_resting-state_2013}
G.~Deco, A.~Ponce-Alvarez, D.~Mantini, G.~L. Romani, P.~Hagmann, M.~Corbetta,
  Resting-state functional connectivity emerges from structurally and
  dynamically shaped slow linear fluctuations, The Journal of Neuroscience: The
  Official Journal of the Society for Neuroscience 33~(27) (2013) 11239--11252.
\newblock \href {http://dx.doi.org/10.1523/JNEUROSCI.1091-13.2013}
  {\path{doi:10.1523/JNEUROSCI.1091-13.2013}}.

\bibitem{park_structural_2013}
H.-J. Park, K.~Friston,
  \href{http://science.sciencemag.org/content/342/6158/1238411}{Structural and
  {Functional} {Brain} {Networks}: {From} {Connections} to {Cognition}},
  Science 342~(6158) (2013) 1238411.
\newblock \href {http://dx.doi.org/10.1126/science.1238411}
  {\path{doi:10.1126/science.1238411}}.
\newline\urlprefix\url{http://science.sciencemag.org/content/342/6158/1238411}

\bibitem{fornito_genetic_2011}
A.~Fornito, A.~Zalesky, D.~S. Bassett, D.~Meunier, I.~Ellison-Wright,
  M.~Yücel, S.~J. Wood, K.~Shaw, J.~O'Connor, D.~Nertney, B.~J. Mowry,
  C.~Pantelis, E.~T. Bullmore, Genetic influences on cost-efficient
  organization of human cortical functional networks, The Journal of
  Neuroscience: The Official Journal of the Society for Neuroscience 31~(9)
  (2011) 3261--3270.
\newblock \href {http://dx.doi.org/10.1523/JNEUROSCI.4858-10.2011}
  {\path{doi:10.1523/JNEUROSCI.4858-10.2011}}.

\bibitem{achard_hubs_2012}
S.~Achard, C.~Delon-Martin, P.~E. Vértes, F.~Renard, M.~Schenck, F.~Schneider,
  C.~Heinrich, S.~Kremer, E.~T. Bullmore,
  \href{http://www.pnas.org/content/109/50/20608}{Hubs of brain functional
  networks are radically reorganized in comatose patients}, Proceedings of the
  National Academy of Sciences 109~(50) (2012) 20608--20613.
\newblock \href {http://dx.doi.org/10.1073/pnas.1208933109}
  {\path{doi:10.1073/pnas.1208933109}}.
\newline\urlprefix\url{http://www.pnas.org/content/109/50/20608}

\bibitem{chennu_spectral_2014}
S.~Chennu, P.~Finoia, E.~Kamau, J.~Allanson, G.~B. Williams, M.~M. Monti,
  V.~Noreika, A.~Arnatkeviciute, A.~Canales-Johnson, F.~Olivares,
  D.~Cabezas-Soto, D.~K. Menon, J.~D. Pickard, A.~M. Owen, T.~A. Bekinschtein,
  \href{http://journals.plos.org/ploscompbiol/article?id=10.1371/journal.pcbi.1003887}{Spectral
  {Signatures} of {Reorganised} {Brain} {Networks} in {Disorders} of
  {Consciousness}}, PLOS Computational Biology 10~(10) (2014) e1003887.
\newblock \href {http://dx.doi.org/10.1371/journal.pcbi.1003887}
  {\path{doi:10.1371/journal.pcbi.1003887}}.
\newline\urlprefix\url{http://journals.plos.org/ploscompbiol/article?id=10.1371/journal.pcbi.1003887}

\bibitem{tijms_alzheimers_2013}
B.~M. Tijms, A.~M. Wink, W.~de~Haan, W.~M. van~der Flier, C.~J. Stam,
  P.~Scheltens, F.~Barkhof,
  \href{http://www.sciencedirect.com/science/article/pii/S0197458013000997}{Alzheimer's
  disease: connecting findings from graph theoretical studies of brain
  networks}, Neurobiology of Aging 34~(8) (2013) 2023--2036.
\newblock \href {http://dx.doi.org/10.1016/j.neurobiolaging.2013.02.020}
  {\path{doi:10.1016/j.neurobiolaging.2013.02.020}}.
\newline\urlprefix\url{http://www.sciencedirect.com/science/article/pii/S0197458013000997}

\bibitem{grefkes_reorganization_2011}
C.~Grefkes, G.~R. Fink,
  \href{http://www.ncbi.nlm.nih.gov/pmc/articles/PMC3097886/}{Reorganization of
  cerebral networks after stroke: new insights from neuroimaging with
  connectivity approaches}, Brain 134~(5) (2011) 1264--1276.
\newblock \href {http://dx.doi.org/10.1093/brain/awr033}
  {\path{doi:10.1093/brain/awr033}}.
\newline\urlprefix\url{http://www.ncbi.nlm.nih.gov/pmc/articles/PMC3097886/}

\bibitem{lynall_functional_2010}
M.-E. Lynall, D.~S. Bassett, R.~Kerwin, P.~J. McKenna, M.~Kitzbichler,
  U.~Muller, E.~Bullmore,
  \href{http://www.jneurosci.org/content/30/28/9477}{Functional {Connectivity}
  and {Brain} {Networks} in {Schizophrenia}}, Journal of Neuroscience 30~(28)
  (2010) 9477--9487.
\newblock \href {http://dx.doi.org/10.1523/JNEUROSCI.0333-10.2010}
  {\path{doi:10.1523/JNEUROSCI.0333-10.2010}}.
\newline\urlprefix\url{http://www.jneurosci.org/content/30/28/9477}

\bibitem{stam_functional_2004}
C.~Stam,
  \href{http://linkinghub.elsevier.com/retrieve/pii/S0304394003012722}{Functional
  connectivity patterns of human magnetoencephalographic recordings: a
  ‘small-world’ network?}, Neuroscience Letters 355~(1-2) (2004) 25--28.
\newblock \href {http://dx.doi.org/10.1016/j.neulet.2003.10.063}
  {\path{doi:10.1016/j.neulet.2003.10.063}}.
\newline\urlprefix\url{http://linkinghub.elsevier.com/retrieve/pii/S0304394003012722}

\bibitem{rubinov_complex_2010}
M.~Rubinov, O.~Sporns,
  \href{http://www.ncbi.nlm.nih.gov/pubmed/19819337}{Complex network measures
  of brain connectivity: uses and interpretations.}, NeuroImage 52~(3) (2010)
  1059--69.
\newblock \href {http://dx.doi.org/10.1016/j.neuroimage.2009.10.003}
  {\path{doi:10.1016/j.neuroimage.2009.10.003}}.
\newline\urlprefix\url{http://www.ncbi.nlm.nih.gov/pubmed/19819337}

\bibitem{stam_organization_2012}
C.~J. Stam, E.~C.~W. van Straaten,
  \href{http://dx.doi.org/10.1016/j.clinph.2012.01.011}{The organization of
  physiological brain networks}, Clinical Neurophysiology 123~(6) (2012)
  1067--1087.
\newblock \href {http://dx.doi.org/10.1016/j.clinph.2012.01.011}
  {\path{doi:10.1016/j.clinph.2012.01.011}}.
\newline\urlprefix\url{http://dx.doi.org/10.1016/j.clinph.2012.01.011}

\bibitem{tumminello_tool_2005}
M.~Tumminello, T.~Aste, T.~D. Matteo, R.~N. Mantegna,
  \href{http://www.pnas.org/content/102/30/10421}{A tool for filtering
  information in complex systems}, Proceedings of the National Academy of
  Sciences of the United States of America 102~(30) (2005) 10421--10426.
\newblock \href {http://dx.doi.org/10.1073/pnas.0500298102}
  {\path{doi:10.1073/pnas.0500298102}}.
\newline\urlprefix\url{http://www.pnas.org/content/102/30/10421}

\bibitem{vidal_interactome_2011}
M.~Vidal, M.~Cusick, A.-L. Barabási,
  \href{http://www.sciencedirect.com/science/article/pii/S0092867411001309}{Interactome
  {Networks} and {Human} {Disease}}, Cell 144~(6) (2011) 986--998.
\newblock \href {http://dx.doi.org/10.1016/j.cell.2011.02.016}
  {\path{doi:10.1016/j.cell.2011.02.016}}.
\newline\urlprefix\url{http://www.sciencedirect.com/science/article/pii/S0092867411001309}

\bibitem{de_vico_fallani_graph_2014}
F.~De~Vico~Fallani, J.~Richiardi, M.~Chavez, S.~Achard, P.~T. R.~S. B, Graph
  analysis of functional brain networks : practical issues in translational
  neuroscience {Graph} analysis of functional brain networks : practical issues
  in translational neuroscience~(September).

\bibitem{goldenberg_survey_2009}
A.~Goldenberg, A.~X. Zheng, S.~E. Fienberg, E.~M. Airoldi,
  \href{http://arxiv.org/abs/0912.5410}{A survey of statistical network
  models}, arXiv:0912.5410 [physics, q-bio, stat]ArXiv: 0912.5410.
\newline\urlprefix\url{http://arxiv.org/abs/0912.5410}

\bibitem{milo_network_2002}
R.~Milo, S.~Shen-Orr, S.~Itzkovitz, N.~Kashtan, D.~Chklovskii, U.~Alon,
  \href{http://science.sciencemag.org/content/298/5594/824}{Network {Motifs}:
  {Simple} {Building} {Blocks} of {Complex} {Networks}}, Science 298~(5594)
  (2002) 824--827.
\newblock \href {http://dx.doi.org/10.1126/science.298.5594.824}
  {\path{doi:10.1126/science.298.5594.824}}.
\newline\urlprefix\url{http://science.sciencemag.org/content/298/5594/824}

\bibitem{garlaschelli_patterns_2004}
D.~Garlaschelli, M.~I. Loffredo,
  \href{http://arxiv.org/abs/cond-mat/0404521}{Patterns of link reciprocity in
  directed networks}, Physical Review Letters 93~(26), arXiv: cond-mat/0404521.
\newblock \href {http://dx.doi.org/10.1103/PhysRevLett.93.268701}
  {\path{doi:10.1103/PhysRevLett.93.268701}}.
\newline\urlprefix\url{http://arxiv.org/abs/cond-mat/0404521}

\bibitem{humphries_network_2008}
M.~D. Humphries, K.~Gurney,
  \href{http://journals.plos.org/plosone/article?id=10.1371/journal.pone.0002051}{Network
  ‘{Small}-{World}-{Ness}’: {A} {Quantitative} {Method} for {Determining}
  {Canonical} {Network} {Equivalence}}, PLOS ONE 3~(4) (2008) e0002051.
\newblock \href {http://dx.doi.org/10.1371/journal.pone.0002051}
  {\path{doi:10.1371/journal.pone.0002051}}.
\newline\urlprefix\url{http://journals.plos.org/plosone/article?id=10.1371/journal.pone.0002051}

\bibitem{barabasi_evolution_2002}
A.-L. Barabasi, H.~Jeong, Z.~Néda, E.~Ravasz, A.~Schubert, T.~Vicsek,
  \href{http://www.sciencedirect.com/science/article/pii/S0378437102007367}{Evolution
  of the social network of scientific collaborations}, Physica A: Statistical
  mechanics and its applications 311~(3) (2002) 590--614.
\newline\urlprefix\url{http://www.sciencedirect.com/science/article/pii/S0378437102007367}

\bibitem{newman_random_2002}
M.~E.~J. Newman, D.~J. Watts, S.~H. Strogatz,
  \href{http://www.pnas.org/content/99/suppl_1/2566}{Random graph models of
  social networks}, Proceedings of the National Academy of Sciences 99~(suppl
  1) (2002) 2566--2572.
\newblock \href {http://dx.doi.org/10.1073/pnas.012582999}
  {\path{doi:10.1073/pnas.012582999}}.
\newline\urlprefix\url{http://www.pnas.org/content/99/suppl_1/2566}

\bibitem{barthelemy_spatial_2011}
M.~Barthélemy,
  \href{http://www.sciencedirect.com/science/article/pii/S037015731000308X}{Spatial
  networks}, Physics Reports 499~(1–3) (2011) 1--101.
\newblock \href {http://dx.doi.org/10.1016/j.physrep.2010.11.002}
  {\path{doi:10.1016/j.physrep.2010.11.002}}.
\newline\urlprefix\url{http://www.sciencedirect.com/science/article/pii/S037015731000308X}

\bibitem{vertes_simple_2012}
P.~E. Vertes, A.~F. Alexander-Bloch, N.~Gogtay, J.~N. Giedd, J.~L. Rapoport,
  E.~T. Bullmore, \href{http://www.pnas.org/content/109/15/5868}{Simple models
  of human brain functional networks}, Proceedings of the National Academy of
  Sciences 109~(15) (2012) 5868--5873.
\newblock \href {http://dx.doi.org/10.1073/pnas.1111738109}
  {\path{doi:10.1073/pnas.1111738109}}.
\newline\urlprefix\url{http://www.pnas.org/content/109/15/5868}

\bibitem{betzel_generative_2016}
R.~F. Betzel, A.~Avena-Koenigsberger, J.~Goñi, Y.~He, M.~A. de~Reus,
  A.~Griffa, P.~E. Vértes, B.~Mišic, J.-P. Thiran, P.~Hagmann, M.~van~den
  Heuvel, X.-N. Zuo, E.~T. Bullmore, O.~Sporns,
  \href{http://www.sciencedirect.com/science/article/pii/S1053811915008563}{Generative
  models of the human connectome}, NeuroImage 124, Part A (2016) 1054--1064.
\newblock \href {http://dx.doi.org/10.1016/j.neuroimage.2015.09.041}
  {\path{doi:10.1016/j.neuroimage.2015.09.041}}.
\newline\urlprefix\url{http://www.sciencedirect.com/science/article/pii/S1053811915008563}

\bibitem{frank_markov_1986}
O.~Frank, D.~Strauss,
  \href{http://amstat.tandfonline.com/doi/abs/10.1080/01621459.1986.10478342}{Markov
  graphs}, Journal of the american Statistical … 81~(395) (1986) 832--842.
\newblock \href {http://dx.doi.org/10.2307/2289017}
  {\path{doi:10.2307/2289017}}.
\newline\urlprefix\url{http://amstat.tandfonline.com/doi/abs/10.1080/01621459.1986.10478342}

\bibitem{frank_statistical_1991}
O.~Frank,
  \href{http://onlinelibrary.wiley.com/doi/10.1111/j.1467-9574.1991.tb01310.x/abstract}{Statistical
  analysis of change in networks}, Statistica Neerlandica 45~(3) (1991)
  283--293.
\newblock \href {http://dx.doi.org/10.1111/j.1467-9574.1991.tb01310.x}
  {\path{doi:10.1111/j.1467-9574.1991.tb01310.x}}.
\newline\urlprefix\url{http://onlinelibrary.wiley.com/doi/10.1111/j.1467-9574.1991.tb01310.x/abstract}

\bibitem{wasserman_logit_1996}
S.~Wasserman, P.~Pattison,
  \href{http://link.springer.com/article/10.1007/BF02294547}{Logit models and
  logistic regressions for social networks: {I}. {An} introduction to {Markov}
  graphs andp}, Psychometrika 61~(3) (1996) 401--425.
\newblock \href {http://dx.doi.org/10.1007/BF02294547}
  {\path{doi:10.1007/BF02294547}}.
\newline\urlprefix\url{http://link.springer.com/article/10.1007/BF02294547}

\bibitem{handcock_statistical_2002}
M.~S. Handcock,
  \href{https://books.google.fr/books?hl=en&lr=&id=DOo-hq1-oF0C&oi=fnd&pg=PA191&dq=handcock+2002+statistical&ots=ECbVg8iYB0&sig=ksP5L-y72HPlb4yG6veoIg-9o_M}{Statistical
  models for social networks: {Inference} and degeneracy}, na, 2002.
\newline\urlprefix\url{https://books.google.fr/books?hl=en&lr=&id=DOo-hq1-oF0C&oi=fnd&pg=PA191&dq=handcock+2002+statistical&ots=ECbVg8iYB0&sig=ksP5L-y72HPlb4yG6veoIg-9o_M}

\bibitem{hunter_inference_2006}
D.~R. Hunter, M.~S. Handcock,
  \href{http://amstat.tandfonline.com/doi/abs/10.1198/106186006X133069}{Inference
  in {Curved} {Exponential} {Family} {Models} for {Networks}}, Journal of
  Computational and Graphical Statistics 15~(3) (2006) 565--583.
\newblock \href {http://dx.doi.org/10.1198/106186006X133069}
  {\path{doi:10.1198/106186006X133069}}.
\newline\urlprefix\url{http://amstat.tandfonline.com/doi/abs/10.1198/106186006X133069}

\bibitem{goodreau_advances_2007}
S.~M. Goodreau, Advances in exponential random graph (p*) models applied to a
  large social network, Social Networks 29~(39) (2007) 231--248.
\newblock \href {http://dx.doi.org/10.1016/j.socnet.2006.08.001}
  {\path{doi:10.1016/j.socnet.2006.08.001}}.

\bibitem{hunter_curved_2007}
D.~R. Hunter, Curved exponential family models for social networks, Social
  Networks 29 (2007) 216--230.
\newblock \href {http://dx.doi.org/10.1016/j.socnet.2006.08.005}
  {\path{doi:10.1016/j.socnet.2006.08.005}}.

\bibitem{rinaldo_geometry_2009}
A.~Rinaldo, S.~E. Fienberg, Y.~Zhou,
  \href{http://projecteuclid.org/euclid.ejs/1243343761}{On the geometry of
  discrete exponential families with application to exponential random graph
  models}, Electronic Journal of Statistics 3 (2009) 446--484.
\newblock \href {http://dx.doi.org/10.1214/08-EJS350}
  {\path{doi:10.1214/08-EJS350}}.
\newline\urlprefix\url{http://projecteuclid.org/euclid.ejs/1243343761}

\bibitem{robins_closure_2009}
G.~Robins, P.~Pattison, P.~Wang, Closure, connectivity and degree
  distributions: {Exponential} random graph (p*) models for directed social
  networks, Social Networks 31 (2009) 105--117.
\newblock \href {http://dx.doi.org/10.1016/j.socnet.2008.10.006}
  {\path{doi:10.1016/j.socnet.2008.10.006}}.

\bibitem{goodreau_birds_2009}
S.~M. Goodreau, J.~a. Kitts, M.~Morris, Birds of a feather, or friend of a
  friend? {Using} exponential random graph models to investigate adolescent
  social networks., Demography 46~(1) (2009) 103--125.
\newblock \href {http://dx.doi.org/10.1353/dem.0.0045}
  {\path{doi:10.1353/dem.0.0045}}.

\bibitem{wang_exponential_2013}
P.~Wang, P.~Pattison, G.~Robins,
  \href{http://dx.doi.org/10.1016/j.socnet.2011.12.004}{Exponential random
  graph model specifications for bipartite networks-{A} dependence hierarchy},
  Social Networks 35~(2) (2013) 211--222.
\newblock \href {http://dx.doi.org/10.1016/j.socnet.2011.12.004}
  {\path{doi:10.1016/j.socnet.2011.12.004}}.
\newline\urlprefix\url{http://dx.doi.org/10.1016/j.socnet.2011.12.004}

\bibitem{niekamp_sexual_2013}
A.~M. Niekamp, L.~a.~G. Mercken, C.~J. P.~a. Hoebe, N.~H. T.~M.
  Dukers-Muijrers, \href{http://dx.doi.org/10.1016/j.socnet.2013.02.006}{A
  sexual affiliation network of swingers, heterosexuals practicing risk
  behaviours that potentiate the spread of sexually transmitted infections: {A}
  two-mode approach}, Social Networks 35~(2) (2013) 223--236.
\newblock \href {http://dx.doi.org/10.1016/j.socnet.2013.02.006}
  {\path{doi:10.1016/j.socnet.2013.02.006}}.
\newline\urlprefix\url{http://dx.doi.org/10.1016/j.socnet.2013.02.006}

\bibitem{simpson_exponential_2011}
S.~L. Simpson, S.~Hayasaka, P.~J. Laurienti,
  \href{http://www.pubmedcentral.nih.gov/articlerender.fcgi?artid=3102079&tool=pmcentrez&rendertype=abstract}{Exponential
  random graph modeling for complex brain networks.}, PloS one 6~(5) (2011)
  e20039--e20039.
\newblock \href {http://dx.doi.org/10.1371/journal.pone.0020039}
  {\path{doi:10.1371/journal.pone.0020039}}.
\newline\urlprefix\url{http://www.pubmedcentral.nih.gov/articlerender.fcgi?artid=3102079&tool=pmcentrez&rendertype=abstract}

\bibitem{sinke_bayesian_2016}
M.~R.~T. Sinke, R.~M. Dijkhuizen, A.~Caimo, C.~J. Stam, W.~M. Otte,
  \href{http://www.sciencedirect.com/science/article/pii/S1053811916301069}{Bayesian
  exponential random graph modeling of whole-brain structural networks across
  lifespan}, NeuroImage 135 (2016) 79--91.
\newblock \href {http://dx.doi.org/10.1016/j.neuroimage.2016.04.066}
  {\path{doi:10.1016/j.neuroimage.2016.04.066}}.
\newline\urlprefix\url{http://www.sciencedirect.com/science/article/pii/S1053811916301069}

\bibitem{goldberger_physiobank_2000}
A.~L. Goldberger, L.~A.~N. Amaral, L.~Glass, J.~M. Hausdorff, P.~C. Ivanov,
  R.~G. Mark, J.~E. Mietus, G.~B. Moody, C.-K. Peng, H.~E. Stanley,
  \href{http://circ.ahajournals.org/content/101/23/e215}{{PhysioBank},
  {PhysioToolkit}, and {PhysioNet}}, Circulation 101~(23) (2000) e215--e220.
\newblock \href {http://dx.doi.org/10.1161/01.CIR.101.23.e215}
  {\path{doi:10.1161/01.CIR.101.23.e215}}.
\newline\urlprefix\url{http://circ.ahajournals.org/content/101/23/e215}

\bibitem{schalk_bci2000:_2004}
G.~Schalk, D.~J. McFarland, T.~Hinterberger, N.~Birbaumer, J.~R. Wolpaw,
  {BCI}2000: a general-purpose brain-computer interface ({BCI}) system, IEEE
  Transactions on Biomedical Engineering 51~(6) (2004) 1034--1043.
\newblock \href {http://dx.doi.org/10.1109/TBME.2004.827072}
  {\path{doi:10.1109/TBME.2004.827072}}.

\bibitem{carter_coherence_1987}
G.~C. Carter, Coherence and time delay estimation, Proceedings of the IEEE
  75~(2) (1987) 236--255.
\newblock \href {http://dx.doi.org/10.1109/PROC.1987.13723}
  {\path{doi:10.1109/PROC.1987.13723}}.

\bibitem{de_vico_fallani_topological_2017}
F.~De~Vico~Fallani, V.~Latora, M.~Chavez,
  \href{http://journals.plos.org/ploscompbiol/article?id=10.1371/journal.pcbi.1005305}{A
  {Topological} {Criterion} for {Filtering} {Information} in {Complex} {Brain}
  {Networks}}, PLOS Computational Biology 13~(1) (2017) e1005305.
\newblock \href {http://dx.doi.org/10.1371/journal.pcbi.1005305}
  {\path{doi:10.1371/journal.pcbi.1005305}}.
\newline\urlprefix\url{http://journals.plos.org/ploscompbiol/article?id=10.1371/journal.pcbi.1005305}

\bibitem{tononi_measure_1994}
G.~Tononi, O.~Sporns, G.~M. Edelman, A measure for brain complexity: relating
  functional segregation and integration in the nervous system., Proceedings of
  the National Academy of Sciences of the United States of America 91~(May)
  (1994) 5033--5037.
\newblock \href {http://dx.doi.org/10.1073/pnas.91.11.5033}
  {\path{doi:10.1073/pnas.91.11.5033}}.

\bibitem{bassett_small-world_2006}
D.~S. Bassett, E.~Bullmore,
  \href{http://nro.sagepub.com/content/12/6/512}{Small-{World} {Brain}
  {Networks}}, The Neuroscientist 12~(6) (2006) 512--523.
\newblock \href {http://dx.doi.org/10.1177/1073858406293182}
  {\path{doi:10.1177/1073858406293182}}.
\newline\urlprefix\url{http://nro.sagepub.com/content/12/6/512}

\bibitem{friston_functional_2011}
K.~J. Friston, \href{http://www.ncbi.nlm.nih.gov/pubmed/22432952}{Functional
  and effective connectivity: a review.}, Brain connectivity 1~(1) (2011)
  13--36.
\newblock \href {http://dx.doi.org/10.1089/brain.2011.0008}
  {\path{doi:10.1089/brain.2011.0008}}.
\newline\urlprefix\url{http://www.ncbi.nlm.nih.gov/pubmed/22432952}

\bibitem{watts_collective_1998}
D.~J. Watts, S.~H. Strogatz,
  \href{http://www.nature.com/nature/journal/v393/n6684/abs/393440a0.html?cookies=accepted}{Collective
  dynamics of ‘small-world’ networks}, Nature 393~(6684) (1998) 440--442.
\newblock \href {http://dx.doi.org/10.1038/30918} {\path{doi:10.1038/30918}}.
\newline\urlprefix\url{http://www.nature.com/nature/journal/v393/n6684/abs/393440a0.html?cookies=accepted}

\bibitem{latora_efficient_2001}
V.~Latora, M.~Marchiori, Efficient behavior of small-world networks., Physical
  review letters 87 (2001) 198701--198701.
\newblock \href {http://dx.doi.org/10.1103/PhysRevLett.87.198701}
  {\path{doi:10.1103/PhysRevLett.87.198701}}.

\bibitem{pons_computing_2005}
P.~Pons, M.~Latapy,
  \href{http://link.springer.com/chapter/10.1007/11569596_31}{Computing
  {Communities} in {Large} {Networks} {Using} {Random} {Walks}}, in: Computer
  and {Information} {Sciences} - {ISCIS} 2005, Springer, Berlin, Heidelberg,
  2005, pp. 284--293, dOI: 10.1007/11569596\_31.
\newline\urlprefix\url{http://link.springer.com/chapter/10.1007/11569596_31}

\bibitem{newman_modularity_2006}
M.~E.~J. Newman, \href{http://www.pnas.org/content/103/23/8577}{Modularity and
  community structure in networks}, Proceedings of the National Academy of
  Sciences 103~(23) (2006) 8577--8582.
\newblock \href {http://dx.doi.org/10.1073/pnas.0601602103}
  {\path{doi:10.1073/pnas.0601602103}}.
\newline\urlprefix\url{http://www.pnas.org/content/103/23/8577}

\bibitem{newman_networks:_2010}
M.~Newman, Networks: {An} {Introduction}, Oxford University Press, 2010.

\bibitem{robins_introduction_2007}
G.~Robins, P.~Pattison, Y.~Kalish, D.~Lusher, An introduction to exponential
  random graph (p *) models for social networks, Social Networks 29 (2007)
  173--191.
\newblock \href {http://dx.doi.org/10.1016/j.socnet.2006.08.002}
  {\path{doi:10.1016/j.socnet.2006.08.002}}.

\bibitem{snijders_markov_2002}
T.~A.~B. Snijders, Markov {Chain} {Monte} {Carlo} {Estimation} of {Exponential}
  {Random} {Graph} {Models}.

\bibitem{amaral_classes_2000}
L.~a.~N. Amaral, A.~Scala, M.~Barthélémy, H.~E. Stanley,
  \href{http://www.pnas.org/content/97/21/11149}{Classes of small-world
  networks}, Proceedings of the National Academy of Sciences 97~(21) (2000)
  11149--11152.
\newblock \href {http://dx.doi.org/10.1073/pnas.200327197}
  {\path{doi:10.1073/pnas.200327197}}.
\newline\urlprefix\url{http://www.pnas.org/content/97/21/11149}

\bibitem{wang_complex_2003}
X.~F. Wang, G.~Chen, Complex networks: small-world, scale-free and beyond, IEEE
  Circuits and Systems Magazine 3~(1) (2003) 6--20.
\newblock \href {http://dx.doi.org/10.1109/MCAS.2003.1228503}
  {\path{doi:10.1109/MCAS.2003.1228503}}.

\bibitem{snijders_new_2006}
P.~P. Snijders, Robins GL \& Handcock~MS,
  \href{http://onlinelibrary.wiley.com/doi/10.1111/j.1467-9531.2006.00176.x/abstract}{New
  specifications for exponential random graph models} (2006) 99--153\href
  {http://dx.doi.org/10.1111/j.1467-9531.2006.00176.x}
  {\path{doi:10.1111/j.1467-9531.2006.00176.x}}.
\newline\urlprefix\url{http://onlinelibrary.wiley.com/doi/10.1111/j.1467-9531.2006.00176.x/abstract}

\bibitem{robins_recent_2007}
G.~Robins, T.~Snijders, P.~Wang, M.~Handcock, P.~Pattison, Recent developments
  in exponential random graph (p*) models for social networks, Social Networks
  29 (2007) 192--215.
\newblock \href {http://dx.doi.org/10.1016/j.socnet.2006.08.003}
  {\path{doi:10.1016/j.socnet.2006.08.003}}.

\bibitem{lusher_exponential_2012}
D.~Lusher, J.~Koskinen, G.~Robins, Exponential {Random} {Graph} {Models} for
  {Social} {Networks}: {Theory}, {Methods}, and {Applications}, Cambridge
  University Press, 2012, google-Books-ID: gyKypohCjDcC.

\bibitem{van_duijn_framework_2009}
M.~A.~J. van Duijn, K.~J. Gile, M.~S. Handcock,
  \href{http://www.sciencedirect.com/science/article/pii/S0378873308000543}{A
  framework for the comparison of maximum pseudo-likelihood and maximum
  likelihood estimation of exponential family random graph models}, Social
  Networks 31~(1) (2009) 52--62.
\newblock \href {http://dx.doi.org/10.1016/j.socnet.2008.10.003}
  {\path{doi:10.1016/j.socnet.2008.10.003}}.
\newline\urlprefix\url{http://www.sciencedirect.com/science/article/pii/S0378873308000543}

\bibitem{akaike_information_1998}
H.~Akaike,
  \href{http://link.springer.com/chapter/10.1007/978-1-4612-1694-0_15}{Information
  {Theory} and an {Extension} of the {Maximum} {Likelihood} {Principle}}, in:
  E.~Parzen, K.~Tanabe, G.~Kitagawa (Eds.), Selected {Papers} of {Hirotugu}
  {Akaike}, Springer {Series} in {Statistics}, Springer New York, 1998, pp.
  199--213, dOI: 10.1007/978-1-4612-1694-0\_15.
\newline\urlprefix\url{http://link.springer.com/chapter/10.1007/978-1-4612-1694-0_15}

\bibitem{meila_comparing_2007}
M.~Meilă,
  \href{https://www.sciencedirect.com/science/article/pii/S0047259X06002016}{Comparing
  clusterings—an information based distance}, Journal of Multivariate
  Analysis 98~(5) (2007) 873--895.
\newblock \href {http://dx.doi.org/10.1016/j.jmva.2006.11.013}
  {\path{doi:10.1016/j.jmva.2006.11.013}}.
\newline\urlprefix\url{https://www.sciencedirect.com/science/article/pii/S0047259X06002016}

\bibitem{guimera_network_2013}
R.~Guimerà, M.~Sales-Pardo,
  \href{http://journals.plos.org/ploscompbiol/article?id=10.1371/journal.pcbi.1003374}{A
  {Network} {Inference} {Method} for {Large}-{Scale} {Unsupervised}
  {Identification} of {Novel} {Drug}-{Drug} {Interactions}}, PLOS Computational
  Biology 9~(12) (2013) e1003374.
\newblock \href {http://dx.doi.org/10.1371/journal.pcbi.1003374}
  {\path{doi:10.1371/journal.pcbi.1003374}}.
\newline\urlprefix\url{http://journals.plos.org/ploscompbiol/article?id=10.1371/journal.pcbi.1003374}

\bibitem{martin_structural_2016}
T.~Martin, B.~Ball, M.~E.~J. Newman,
  \href{http://arxiv.org/abs/1506.05490}{Structural inference for uncertain
  networks}, Physical Review E 93~(1), arXiv: 1506.05490.
\newblock \href {http://dx.doi.org/10.1103/PhysRevE.93.012306}
  {\path{doi:10.1103/PhysRevE.93.012306}}.
\newline\urlprefix\url{http://arxiv.org/abs/1506.05490}

\bibitem{hric_network_2016}
D.~Hric, T.~P. Peixoto, S.~Fortunato,
  \href{http://arxiv.org/abs/1604.00255}{Network structure, metadata and the
  prediction of missing nodes and annotations}, Physical Review X 6~(3), arXiv:
  1604.00255.
\newblock \href {http://dx.doi.org/10.1103/PhysRevX.6.031038}
  {\path{doi:10.1103/PhysRevX.6.031038}}.
\newline\urlprefix\url{http://arxiv.org/abs/1604.00255}

\bibitem{craddock_imaging_2013}
R.~C. Craddock, S.~Jbabdi, C.-G. Yan, J.~T. Vogelstein, F.~X. Castellanos,
  A.~Di~Martino, C.~Kelly, K.~Heberlein, S.~Colcombe, M.~P. Milham,
  \href{http://www.nature.com/nmeth/journal/v10/n6/abs/nmeth.2482.html?cookies=accepted}{Imaging
  human connectomes at the macroscale}, Nature Methods 10~(6) (2013) 524--539.
\newblock \href {http://dx.doi.org/10.1038/nmeth.2482}
  {\path{doi:10.1038/nmeth.2482}}.
\newline\urlprefix\url{http://www.nature.com/nmeth/journal/v10/n6/abs/nmeth.2482.html?cookies=accepted}

\bibitem{holland_stochastic_1983}
P.~W. Holland, K.~B. Laskey, S.~Leinhardt,
  \href{http://www.sciencedirect.com/science/article/pii/0378873383900217}{Stochastic
  blockmodels: {First} steps}, Social networks 5~(2) (1983) 109--137.
\newline\urlprefix\url{http://www.sciencedirect.com/science/article/pii/0378873383900217}

\bibitem{barabasi_emergence_1999}
A.-L. Barabasi, R.~Albert,
  \href{http://science.sciencemag.org/content/286/5439/509}{Emergence of
  {Scaling} in {Random} {Networks}}, Science 286~(5439) (1999) 509--512.
\newblock \href {http://dx.doi.org/10.1126/science.286.5439.509}
  {\path{doi:10.1126/science.286.5439.509}}.
\newline\urlprefix\url{http://science.sciencemag.org/content/286/5439/509}

\bibitem{sporns_network_2013}
O.~Sporns,
  \href{http://www.sciencedirect.com/science/article/pii/S0959438812001894}{Network
  attributes for segregation and integration in the human brain}, Current
  Opinion in Neurobiology 23~(2) (2013) 162--171.
\newblock \href {http://dx.doi.org/10.1016/j.conb.2012.11.015}
  {\path{doi:10.1016/j.conb.2012.11.015}}.
\newline\urlprefix\url{http://www.sciencedirect.com/science/article/pii/S0959438812001894}

\bibitem{deco_rethinking_2015}
G.~Deco, G.~Tononi, M.~Boly, M.~L. Kringelbach,
  \href{http://www.nature.com/nrn/journal/v16/n7/abs/nrn3963.html}{Rethinking
  segregation and integration: contributions of whole-brain modelling}, Nature
  Reviews Neuroscience 16~(7) (2015) 430--439.
\newblock \href {http://dx.doi.org/10.1038/nrn3963}
  {\path{doi:10.1038/nrn3963}}.
\newline\urlprefix\url{http://www.nature.com/nrn/journal/v16/n7/abs/nrn3963.html}

\bibitem{karrer_stochastic_2011}
B.~Karrer, M.~E.~J. Newman, \href{http://arxiv.org/abs/1008.3926}{Stochastic
  blockmodels and community structure in networks}, Physical Review E 83~(1),
  arXiv: 1008.3926.
\newblock \href {http://dx.doi.org/10.1103/PhysRevE.83.016107}
  {\path{doi:10.1103/PhysRevE.83.016107}}.
\newline\urlprefix\url{http://arxiv.org/abs/1008.3926}

\bibitem{pavlovic_stochastic_2014}
D.~M. Pavlovic, P.~E. Vértes, E.~T. Bullmore, W.~R. Schafer, T.~E. Nichols,
  \href{http://journals.plos.org/plosone/article?id=10.1371/journal.pone.0097584}{Stochastic
  {Blockmodeling} of the {Modules} and {Core} of the {Caenorhabditis} elegans
  {Connectome}}, PLOS ONE 9~(7) (2014) e97584.
\newblock \href {http://dx.doi.org/10.1371/journal.pone.0097584}
  {\path{doi:10.1371/journal.pone.0097584}}.
\newline\urlprefix\url{http://journals.plos.org/plosone/article?id=10.1371/journal.pone.0097584}

\bibitem{barry_eeg_2007}
R.~J. Barry, A.~R. Clarke, S.~J. Johnstone, C.~A. Magee, J.~A. Rushby,
  \href{http://www.sciencedirect.com/science/article/pii/S1388245707004002}{{EEG}
  differences between eyes-closed and eyes-open resting conditions}, Clinical
  Neurophysiology 118~(12) (2007) 2765--2773.
\newblock \href {http://dx.doi.org/10.1016/j.clinph.2007.07.028}
  {\path{doi:10.1016/j.clinph.2007.07.028}}.
\newline\urlprefix\url{http://www.sciencedirect.com/science/article/pii/S1388245707004002}

\bibitem{yan_spontaneous_2009}
C.~Yan, D.~Liu, Y.~He, Q.~Zou, C.~Zhu, X.~Zuo, X.~Long, Y.~Zang,
  \href{http://journals.plos.org/plosone/article?id=10.1371/journal.pone.0005743}{Spontaneous
  {Brain} {Activity} in the {Default} {Mode} {Network} {Is} {Sensitive} to
  {Different} {Resting}-{State} {Conditions} with {Limited} {Cognitive}
  {Load}}, PLOS ONE 4~(5) (2009) e5743.
\newblock \href {http://dx.doi.org/10.1371/journal.pone.0005743}
  {\path{doi:10.1371/journal.pone.0005743}}.
\newline\urlprefix\url{http://journals.plos.org/plosone/article?id=10.1371/journal.pone.0005743}

\bibitem{marx_eyes_2004}
E.~Marx, A.~Deutschländer, T.~Stephan, M.~Dieterich, M.~Wiesmann, T.~Brandt,
  \href{http://www.sciencedirect.com/science/article/pii/S1053811903007936}{Eyes
  open and eyes closed as rest conditions: impact on brain activation
  patterns}, NeuroImage 21~(4) (2004) 1818--1824.
\newblock \href {http://dx.doi.org/10.1016/j.neuroimage.2003.12.026}
  {\path{doi:10.1016/j.neuroimage.2003.12.026}}.
\newline\urlprefix\url{http://www.sciencedirect.com/science/article/pii/S1053811903007936}

\bibitem{bianciardi_modulation_2009}
M.~Bianciardi, M.~Fukunaga, P.~van Gelderen, S.~G. Horovitz, J.~A. de~Zwart,
  J.~H. Duyn,
  \href{http://www.ncbi.nlm.nih.gov/pmc/articles/PMC2704889/}{Modulation of
  spontaneous {fMRI} activity in human visual cortex by behavioral state},
  NeuroImage 45~(1) (2009) 160--168.
\newblock \href {http://dx.doi.org/10.1016/j.neuroimage.2008.10.034}
  {\path{doi:10.1016/j.neuroimage.2008.10.034}}.
\newline\urlprefix\url{http://www.ncbi.nlm.nih.gov/pmc/articles/PMC2704889/}

\bibitem{xu_different_2014}
P.~Xu, R.~Huang, J.~Wang, N.~T. Van~Dam, T.~Xie, Z.~Dong, C.~Chen, R.~Gu, Y.-F.
  Zang, Y.~He, J.~Fan, Y.-j. Luo,
  \href{http://www.sciencedirect.com/science/article/pii/S1053811914000056}{Different
  topological organization of human brain functional networks with eyes open
  versus eyes closed}, NeuroImage 90 (2014) 246--255.
\newblock \href {http://dx.doi.org/10.1016/j.neuroimage.2013.12.060}
  {\path{doi:10.1016/j.neuroimage.2013.12.060}}.
\newline\urlprefix\url{http://www.sciencedirect.com/science/article/pii/S1053811914000056}

\bibitem{boytsova_eeg_2010}
Y.~A. Boytsova, S.~G. Danko,
  \href{http://link.springer.com/article/10.1134/S0362119710030199}{{EEG}
  differences between resting states with eyes open and closed in darkness},
  Human Physiology 36~(3) (2010) 367--369.
\newblock \href {http://dx.doi.org/10.1134/S0362119710030199}
  {\path{doi:10.1134/S0362119710030199}}.
\newline\urlprefix\url{http://link.springer.com/article/10.1134/S0362119710030199}

\bibitem{srinivasan_eeg_2007}
R.~Srinivasan, W.~R. Winter, J.~Ding, P.~L. Nunez,
  \href{http://www.sciencedirect.com/science/article/pii/S016502700700307X}{{EEG}
  and {MEG} coherence: {Measures} of functional connectivity at distinct
  spatial scales of neocortical dynamics}, Journal of Neuroscience Methods
  166~(1) (2007) 41--52.
\newblock \href {http://dx.doi.org/10.1016/j.jneumeth.2007.06.026}
  {\path{doi:10.1016/j.jneumeth.2007.06.026}}.
\newline\urlprefix\url{http://www.sciencedirect.com/science/article/pii/S016502700700307X}

\bibitem{colclough_how_2016}
G.~L. Colclough, M.~W. Woolrich, P.~K. Tewarie, M.~J. Brookes, A.~J. Quinn,
  S.~M. Smith,
  \href{http://www.sciencedirect.com/science/article/pii/S1053811916301914}{How
  reliable are {MEG} resting-state connectivity metrics?}, NeuroImage 138
  (2016) 284--293.
\newblock \href {http://dx.doi.org/10.1016/j.neuroimage.2016.05.070}
  {\path{doi:10.1016/j.neuroimage.2016.05.070}}.
\newline\urlprefix\url{http://www.sciencedirect.com/science/article/pii/S1053811916301914}

\bibitem{garrison_stability_2015}
K.~A. Garrison, D.~Scheinost, E.~S. Finn, X.~Shen, R.~T. Constable, The
  (in)stability of functional brain network measures across thresholds,
  NeuroImage 118 (2015) 651--661.
\newblock \href {http://dx.doi.org/10.1016/j.neuroimage.2015.05.046}
  {\path{doi:10.1016/j.neuroimage.2015.05.046}}.

\bibitem{vogelstein_fast_2011}
J.~T. Vogelstein, J.~M. Conroy, V.~Lyzinski, L.~J. Podrazik, S.~G. Kratzer,
  E.~T. Harley, D.~E. Fishkind, R.~J. Vogelstein, C.~E. Priebe,
  \href{http://arxiv.org/abs/1112.5507}{Fast {Approximate} {Quadratic}
  {Programming} for {Large} ({Brain}) {Graph} {Matching}}, arXiv:1112.5507 [cs,
  math, q-bio]ArXiv: 1112.5507.
\newline\urlprefix\url{http://arxiv.org/abs/1112.5507}

\end{thebibliography}

%TABLES
\newpage

\begin{table}[!ht]
\centering
\begin{tabular}{ccccccl}    \toprule
Models    & Edges  & $GW_K$  & $GW_E$  & $GW_N$ & $GW_D$   \\\midrule
$M_1$ & $\ast$ & \checkmark & \checkmark & - & -  \\ 
$M_2$ & $\ast$ &- & \checkmark & - & \checkmark   \\
$M_3$ & $\ast$ &-  & - & \checkmark & \checkmark \\
$M_4$ & $\ast$ &-  & \checkmark & \checkmark & - \\
$M_5$ & $\ast$ &-  & \checkmark & - & - \\
$M_6$ & $\ast$ &-  & - & \checkmark & - \\
$M_7$ & $\ast$ & \checkmark  & - & \checkmark & - \\
$M_8$ & $\ast$ & \checkmark  & - & - & \checkmark \\
$M_{9}$ & $\ast$ & \checkmark   & - & - & - \\
$M_{10}$ & $\ast$ & -  & - & - & \checkmark \\
$M_{11}$ & \checkmark &-  & \checkmark & \checkmark & - \\\bottomrule
 \hline
\end{tabular}
\caption{Set of model configurations. Models $M_1$ to $M_{10}$ include at most two of the four considered graph metrics i.e., $GW_K$, $GW_E$, $GW_N$, $GW_D$. The metric "Edges" is fixed and equal to the actual number of edges in the observed brain networks in all the configurations but $M_{11}$ model \newline 
$\ast$ Metrics that are fixed.\newline
\checkmark Metrics that are variable.}
\label{table:testGeo}
\end{table}

\newpage
\begin{table}[!ht]
\centering
\resizebox{\textwidth}{!}{%
\begin{tabular}{lccc|ccc} \toprule
\multicolumn{1}{c}{}            & \multicolumn{3}{c|}{$\theta_1$}                                                                                             & \multicolumn{3}{c}{$\theta_2$}                                                                                        \\\midrule
\multicolumn{1}{l|}{} & \multicolumn{1}{c}{EO}    & \multicolumn{1}{c}{EC}     & \multicolumn{1}{c|}{EO-EC}   & \multicolumn{1}{c}{EO}     & \multicolumn{1}{c}{EC}  & \multicolumn{1}{c}{EO-EC}   \\
\multicolumn{1}{l|}{$theta$}      & 1.528(0.045)                         & 1.531(0.039)       &-0.281(0.7804)          & 1.502(0.169)                          & 1.443(0.159)      &-0.406                (0.690)        \\
\multicolumn{1}{l|}{$alpha$}     & 1.449(0.041)                         & 1.297(0.039)       &3.746(0.0002)           & 1.327(0.123)                          & 1.317(0.532)      &-1.084                (0.347)         \\
\multicolumn{1}{l|}{$beta$}       & 1.487(0.457)                         & 1.326(0.046)       &1.514(0.0009)           & 1.062(0.149)                          & 1.303(0.169)      &-0.890                (0.371)        \\
\multicolumn{1}{l|}{$gamma$}  & 1.552(0.046)                         & 1.509(8.266)       &-0.992(0.8521)           & 0.878(0.125)                          & 1.140(3.002)     &-1.064                 (0.135)         \\\bottomrule 
\hline                        
\end{tabular}%
}
\caption{Statistics for the estimated parameters of the model configuration $M_1$. Median values and standard errors (within parentheses) are reported for the two resting-state conditions $EO$ and $EC$. $T$-values and $p$-values (within parentheses) from non-parametric permutation-based t-tests between $EO$ and $EC$ are shown in the third column of each subsection marked with the headline EO-EC. }
\label{table2}
\end{table}

%FIGURES
\newpage

\newpage

\newpage

\newpage

\newpage

\newpage

%\includepdf[pages={1}]{FigureS1.pdf}
%
%\includepdf[pages={1}]{FigureS2.pdf}
%
%\includepdf[pages={1}]{FigureS3.pdf}
%
%\includepdf[pages={1}]{TableS1.pdf}
%
%\includepdf[pages={1}]{TableS2.pdf}
%
%\includepdf[pages={2}]{TableS3.pdf}
%
%\includepdf[pages={1-4}]{Supp_text.pdf}

\end{document}